\documentclass[
11pt,%
final,%
tightenlines,%
notitlepage,%
twoside,%
onecolumn,%
nobibnotes,%
nofootinbib,%
superscriptaddress,%
noshowpacs,%
noshowydk,%
showdoi,%
centertags,%
maik]%
{revtex4n}

\input jpack_e
\setcaptionmargin{5mm}
\onelinecaptionstrue

\columntovsizetrue
\allowdisplaybreaks

\begin{document}

\newtheorem{teo}{Theorem}
\newtheorem*{teon}{Theorem}
\newtheorem{lem}{Lemma}
\newtheorem*{lemn}{Lemma}
\newtheorem{prp}{Proposition}
\newtheorem*{prpn}{Proposition}
\newtheorem{ass}{Assertion}
\newtheorem*{assn}{Assertion}
\newtheorem{assum}{Assumption}
\newtheorem*{assumn}{Assumption}
\newtheorem{stat}{Statement}
\newtheorem*{statn}{Statement}
\newtheorem{cor}{Corollary}
\newtheorem*{corn}{Corollary}
\newtheorem{hyp}{Hypothesis}
\newtheorem*{hypn}{Hypothesis}
\newtheorem{con}{Conjecture}
\newtheorem*{conn}{Conjecture}
\newtheorem{dfn}{Definition}
\newtheorem*{dfnn}{Definition}
\newtheorem{problem}{Problem}
\newtheorem*{problemn}{Problem}
\newtheorem{notat}{Notation}
\newtheorem*{notatn}{Notation}
\newtheorem{quest}{Question}
\newtheorem*{questn}{Question}

\theorembodyfont{\rm}
\newtheorem{rem}{Remark}
\newtheorem*{remn}{Remark}
\newtheorem{exa}{Example}
\newtheorem*{exan}{Example}
\newtheorem{cas}{Case}
\newtheorem*{casn}{Case}
\newtheorem{claim}{Claim}
\newtheorem*{claimn}{Claim}
\newtheorem{com}{Comment}
\newtheorem*{comn}{Comment}

\theoremheaderfont{\it}
\theorembodyfont{\rm}

\newtheorem{proof}{Proof}
\newtheorem*{proofn}{Proof}

\selectlanguage{english}
\Rubrika{\relax}
\CRubrika{\relax}
\SubRubrika{\relax}
\CSubRubrika{\relax}
%

\def\JournalNumber{0}
\def\JournalVolume{00}
%
%
%
\nameVolumeRus{}
\CnameVolumeRus{}
\nameIssueRus{\No}
\CnameIssueRus{}
\namePartRus{}
\namePagesRus{}
\nameYearShortRus{}
\JournalNameRus{}
\TranslitJournalNameRus{}
\JournalName{Regular and Chaotic Dynamics}
\JournalISSNCode{1560-3547}
\IssuePrice{}
\TransYearOfIssue{0000}
\TransCopyrightYear{2016}%
\OrigYearOfIssue{}
\OrigCopyrightYear{2016}%
\OrigIssueNo{\JournalNumber}
\OrigVolumeNo{\JournalVolume}
\TransVolumeNo{\JournalVolume}
\TransIssueNo{\JournalNumber}
\TransPartNo{}
\SHORTjournalPREFIX{RCD} 
\LONGjournalPREFIX{RegDyn} 
\BatFileName{call make_ps.bat} 
\BatSwitch{3} 
\IssueName{}
\SupplementNumber{}
\PublicationSerialNumberInYear{0}
\PublicationSerialNumberInVolume{0}
\ConditionalIssueDate{"year","month","day","name","type"}
\PagePrefix{}
\JournalISSNonlineCode{}
\JournalISSNCodeRus{}
\JournalISSNonlineCodeRus{}
\VolumeName{}
\IssnoName{none}
\PartnoName{}
\FpageNamepp{}
\FpageNnamep{}
\FpagePrefix{}
\LpageNnamepp{}
\LpageNamep{}
\LpagePrefix{}
\VolumePageNumbering{}
\JournalPubID{}
\FirstJournalPageNumber{}
\LastJournalPageNumber{}
\makeatletter
\def\MAIKlogo{RCD Editorial Office}
\def\maikpraefix{10.0000/S}
\edef\@ContentsHeadLineB{Simultaneous English language translation of the journal is available from \noexpand\MAIKlogo}
\def\Distributed{Distributed worldwide by Springer. }
\def\ArticlePages#1{\relax}
\@ifxundefined\CONT@sw{\@booleantrue\CONT@sw}{}%
\@booleantrue\showPACS@sw%
\@booleantrue\showKEYS@sw %
\@booleantrue\noOrigJournalVersion@sw
\@booleantrue\noOrigVolumeNo@sw
\@booleanfalse\noTransVolumeNo@sw
\makeatother
\input maikdoi %

\beginpaper


\input engnames
\titlerunning{Classical and quantum dynamical manifestations of index-$2$ saddles}
\authorrunning{Priyanka Pandey, Shibabrat Naik, Srihari Keshavamurthy}
\toctitle{Title}
\tocauthor{F.\,S.\,Author}
\title{Classical and quantum dynamical manifestations of index-$2$ saddles: concerted versus sequential 
reaction mechanisms }
\firstaffiliation{
}%
\articleinenglish 
\PublishedInRussianNo
\author{\firstname{Priyanka}~\surname{Pandey}}%
\email[E-mail: ]{priyanka@iitk.ac.in}
\affiliation{
Department of Chemistry, Indian Institute of Technology \\
Kanpur, Uttar Pradesh 208016, India }%
\author{\firstname{Shibabrat}~\surname{Naik}}%
\email[E-mail: ]{s.naik@bristol.ac.uk}
\affiliation{
School of Mathematics, University of Bristol \\
Fry Building, Woodland Road, Bristol BS8 1UG, United Kingdom }%
\author{\firstname{Srihari}~\surname{Keshavamurthy}}%
\email[E-mail: ]{srihari@iitk.ac.in}
\affiliation{
Department of Chemistry, Indian Institute of Technology \\
Kanpur, Uttar Pradesh 208016, India }%

\begin{abstract}
The presence of higher index saddles on a multidimensional potential energy surface is usually assumed to be of little significance in chemical reaction dynamics. Such a viewpoint requires careful reconsideration, thanks to elegant experiments and novel theoretical approaches that have come about in recent years. In this work, we perform a detailed classical and quantum dynamical study of a model two degree of freedom Hamiltonian, which captures the essence of the debate regarding the dominance of a concerted or a stepwise reaction mechanism.  We show that the ultrafast shift of the mechanism from a concerted to a stepwise one is essentially a classical dynamical effect. In addition, due to the classical phase space being a mixture of regular and chaotic dynamics, it is possible to have a rich variety of dynamical behaviour, including a Murrell-Laidler type of mechanism, even at energies sufficiently above that of the index-$2$ saddle. We rationalize the dynamical results using an explicit construction of the classical invariant manifolds in the phase space.
\end{abstract}
\received{10 Sept 2020}
\revised{Month XX, 20XX}
\accepted{Month XX, 20XX}%
\maketitle

\textmakefnmark{0}{)}%


\section{INTRODUCTION}

Transition state theory (TST) is a cornerstone for reaction dynamics~\cite{eyring1935activated}. The simple idea of calculating the rate of a reaction as a flux through a ``bottleneck" region (TS) has proved to be of immense utility in chemistry for nearly a century now. The notion of associating the TS with a saddle point on the multidimensional potential energy surface (PES) has led to the development of several powerful algorithms to determine the so called intrinsic reaction coordinate (IRC) or minimum energy path (MEP) and accurate ab initio force fields in the vicinity of the various TS. Although the IRCs and their connectivity across the high dimensional PES are useful starting points for analyzing the reactions~\cite{maeda2013systematic}, the fact remains that MEPs and IRCs are non-dynamical.  Consequently, depending on the energy and other parameters of interest, significant deviations from IRCs may be observed~\cite{tsutsumi2020visualization}. There is little doubt that the true TS is to be found in the full phase space of the system. This dynamical ``Wignerian" view was brought out beautifully by the early work by Pollak and Pechukas~\cite{pollak_transition_1978,pechukas_classical_1979} and more recently generalized to higher degrees of freedom by Wiggins and coworkers~\cite{waalkens_wigners_2008,uzer2002geometry}.

Although, traditionally TS have been associated with index-1 saddles on the PES, in systems with a large number of degrees of freedom the true (dynamical) TS rarely coincides with the saddle points on the PES. Moreover, in such multidimensional systems there are higher index saddles on the PES that are expected to be dynamically relevant for reactions. Indeed, reactions at sufficiently high energies can occur via  pathways that involve both the index-1 saddles as well as the higher index ones. Competition between the different pathways, apart from complicating the mechanistic understanding, can lead to  unexpected products. For example, in a recent experimental study, Lu et al. found that a short and intense extreme ultraviolet (XUV) excited CO$_{2}$ molecule can result in the production of O$_{2}$ molecules~\cite{lu2014evidence}. Similarly,  the ring opening reaction of cyclobutene to $1,3$-butadiene can occur via a disrotatory pathway, as opposed to the conrotatory pathway predicted based on the Woodward-Hoffman rules, due to the involvement of a index-2 saddle.~\cite{quapp2015embedding,breulet1984conrotatory} It is crucial to note here that the implication of higher index saddles to reactivity does not necessarily come under the ambit of the Murrell-Laidler theorem~\cite{murrell1968symmetries,berry1992limitations,heidrich1986saddle} or the McIver-Stanton rules~\cite{stanton1975group,trindle1975group}. The reason being that these rules are based on the IRC perspective, and hence non-dynamical. Note, however, that even within the IRC perspective it is possible for a index-$1$ saddle  to be linked to a higher index saddle which can then lead to multiple products. An example for this comes from the recent work of Harabuchi et al\cite{harabuchi2016nontotally} resulting in the so called nontotally symmetric trifurcation of a reaction pathway. Similarly, modulating reactivity using external forces  need not be constrained by IRC-based rules and can possibly involve pathways that utilize higher index saddles. Thus in  mechanochemistry~\cite{quapp2018toward,haruta2019force,wollenhaupt2015should,ribas2012covalent,mauguiere2013bond} the energetics of the various index saddles alone is no longer a useful criteria and it is quite feasible for the ``forced" dynamics to prefer pathways that explore the various high index saddle regions.

Recently there have been a number of studies that focus on the dynamical aspects of the higher index saddles. A key motivation is to generalize the dynamical perspective of TST to construct locally recrossing free dividing surfaces. Efforts along these lines have been made by Collins et al. on a model two degrees of freedom potential~\cite{collins2011index}, by Haller et al. on the double ionization of helium in an external electric field~\cite{haller2010transition}, and by Nagahata et al. on the proton exchange reaction involving the H$_{5}^{+}$ moiety~\cite{nagahata2013reactivity}. The latter two examples being three degrees of freedom system.  We note that typically the higher index saddles occur along with the usual index-1 saddles due to topological constraints on the potential energy surface. An early work by Mann and Hase highlighted the role of dynamics in the electrocyclic ring opening reaction of cyclopropane radical to form the allyl radical~\cite{mann2002ab}. More recently, Pradhan and Lourderaj performed extensive ab initio dynamics calculations to emphasize the key role of a index-2 saddle point in the denitrogenation reaction of $1$-pyrazoline~\cite{pradhan2019can}. Interestingly, in the study by Mann and Hase the trajectories were not propagated for a sufficiently long time to ascertain if the reaction was indeed statistical or not. On the other hand, Pradhan and Lourderaj did a much more extensive computation, and established significant deviations from the minimum energy path (MEP). They suggested the possibility of slow intramolecular vibrational energy redistribution (IVR) leading to nonstatistical dynamics. 
One phenomenon where the issue of competing pathways, possibly mediated by higher index saddles, continues to be of interest is that of intramolecular double proton transfer (DPT). A number of studies have addressed various aspects of the mechanism of the DPT phenomenon in a  variety of molecular systems. Of particular interest in such studies is the issue of whether the mechanism of DPT is concerted or stepwise. There are indications that the mechanism is sensitive to a variety of factors including temperature and the extent of coupling between the two local hydrogen stretching modes~\cite{yoshikawa2012quantum}. Suggestions for the two hydrogens to be quantum mechanically entangled ~\cite{smedarchina2018entanglement,fillaux2005quantum} and the role of nuclear quantum effects at temperatures below $100$ K~\cite{litman2019elucidating} have been made in porphycene and other structurally related molecules. Careful and rather detailed NMR experiments on DPT in the azophenine led Rumpel and Limbach~\cite{rumpel1989nmr} to observe the breakdown of the so called ``rule of geometric mean". According to this rule, the rate constants $k_{\rm HH}, k_{\rm HD}$, and $k_{\rm DD}$ for DPT reaction with single and double isotopic substitutions are expected to obey the relation 
\begin{equation}
    k_{\rm HD} = \sqrt{k_{\rm HH} k_{\rm DD}} \implies \frac{k_{\rm HH}}{k_{\rm HD}} = \frac{k_{\rm HD}}{k_{\rm DD}}
    \label{rgm_DPT}
\end{equation}
The significant breakdown of the above relation, observed in azophenine and other systems~\cite{meschede1991dynamic}, implies that the mechanism is stepwise rather than concerted. An elegant analysis of a simple model for the isotope effect in DPT can be found in the article by Albery~\cite{albery1986isotope}. However, the fact that the relation Eq.~\eqref{rgm_DPT} is a consequence of equilibrium and traditional TST based arguments implies that dynamical effects are expected to be important in these class of molecules. 

Despite many studies, there are still doubts as to whether the mechanism of DPT can be purely concerted or stepwise. In fact there are hints 
that the dynamics underlying DPT reactions is fairly complicated and quite distinct from single proton transfer reactions.  For instance,  Accardi et al. in their quantum wavepacket dynamical studies on a model two degree of freedom system showed that mechanisms can switch on an ultrafast timescales~\cite{accardi2010synchronous}. Dynamical studies in full dimensionality, both ab initio~\cite{ushiyama2001successive} and Car-Parinello~\cite{walewski2010car}, also indicate the complexity of the reaction mechanism. Indeed, in the context of Diels-Alder reactions, Houk and coworkers~\cite{black2012dynamics} have argued that the mechanism can be described as ``dynamically concerted" since the time lag between the transfer of the two protons is less than the characteristic vibrational period. The wavepacket dynamical results, according to  Accardi et al., is therefore possibly a ``quantization" of the notion of classical synchronicity proposed by Houk and coworker. We also refer the reader to the work\cite{takeuchipnas2007} of Takeuchi and Tahara for an illuminating discussion of the concerted versus stepwise debate in the context of DPT in the $7$-azaindole dimer.

From the above discussion it is clear that the mechanistic implications of higher index saddles even in  gas phase reactions require careful dynamical considerations. In particular, the nature of IVR and consequently the possibility of emergence of new pathways and mechanisms due to the coupling between modes deserves close attention. An  important issue  is whether the classically predicted dynamical pathways and preferences will be respected by the quantum dynamics. For instance, can quantum tunneling lead to the switching of a stepwise isomerization pathway to a concerted one?  Moreover, to date, the signatures of the index-2 saddles on the quantum eigenstates and dynamics has not been explained clearly enough. We note that although the work of Accardi et al.\cite{accardi2010synchronous} is suggestive of a quantum manifestation of the classical synchronicity, a confirmation by performing full dimensional quantum dynamics on the molecules studied by Houk and coworkers\cite{black2012dynamics} is not practical yet. In this article, we perform extensive classical and quantum dynamical studies on the model system\cite{collins2011index} proposed by Collins et al., which is closely related to the earlier models of DPT, to address the above issues. We demonstrate a striking classical-quantum dynamical correspondence and argue that much of the mechanism of DPT at high energies can be understood from a classical phase space perspective.

\section{MODEL HAMILTONIAN}
To study the dynamics in the proximity of the index-2 saddle, we consider a classical two degrees of freedom Hamiltonian of the form 
\begin{equation}
  H(\mathbf{P},\mathbf{Q})=H_0(\mathbf{P},\mathbf{Q})+V_{coup}(\mathbf{Q})  
\label{eq:hamiltonian}
\end{equation}
where $(\mathbf{P},\mathbf{Q})$ are momentum and position variables. The zeroth-order Hamiltonian term is given by  
\begin{equation}
  H_0(\mathbf{P},\mathbf{Q})=\sum_{j=1,2} \left[\frac{1}{2m_{\rm H}} P_{j}^{2} + V_{j}(Q_{j}) \right] 
  \label{eq:zeroth_hamiltonian}
\end{equation}
 with the symmetric double-well potential term
\begin{equation}
    V_{j}(Q_{j}) = -a_{j}Q_{j}^{2} + b_{j} Q_{j}^{4}
    \label{eq:mode_potential}
\end{equation}
The coupling term   
\begin{equation}
 V_{coup}(\mathbf{Q}) = \gamma Q_{1}^{2} Q_{2}^{2}  
 \label{eq:coupling_potential}
\end{equation}  
preserves the symmetry and $\gamma$ is the coupling strength.

The model Hamiltonian in Eq.\eqref{eq:hamiltonian} has been used to describe the double proton transfer (DPT) in a number of studies. In particular, the coordinates $(Q_1,Q_2)$ are essentially the collective coordinates introduced in a detailed study of DPT by Smedarchina et al~\cite{smedarchina2007correlated,smedarchina2008mechanisms}. The mode-mode coupling in Eq.~(\ref{eq:coupling_potential}) may appear to be a special choice. However, as elucidated by Smedarchina et al, the  coupling term $\gamma Q_{1}^{2} Q_{2}^{2}$ in the collective coordinates accounts for couplings up to fourth order in the original local proton transfer coordinates. This can be seen by transforming the potential in terms of the local proton transfer coordinates $q_{1,2} \equiv (Q_1 \pm Q_2)/2$, resulting in the potential $V({\bf q}) = V_0({\bf q}) + V_{\rm coup}(\bf q)$
with the zeroth-order part
\begin{equation}
    V_0({\bf q}) = -\left(\frac{a_1 + a_2}{4}\right) (q_{1}^{2} + q_{2}^{2}) + \left(\frac{b_1 + b_2}{16}\right) (q_{1}^{4} + q_{2}^{4})
\end{equation}
and the coupling term
\begin{equation}
    V_{\rm coup}({\bf q}) = -\left(\frac{a_1 + a_2}{2}\right) q_1 q_2 + \left(\frac{b_1 - b_2}{4}\right) (q_{1}^{3} q_{2} + q_{1} q_{2}^{3}) + \left(\frac{3(b_1 + b_2)}{8}\right) q_{1}^{2} q_{2}^{2}
    \label{eq:couplocalcoords}
\end{equation}
Thus, the potential in the collective coordinates $V(\bf Q)$ with a single coupling term $\gamma Q_{1}^{2} Q_{2}^{2}$ accounts for the various couplings in $V({\bf q})$. However, note that in the context of DPT the last term in Eq.~(\ref{eq:couplocalcoords}) is not allowed since symmetry considerations demand that the couplings must be symmetric in the two coordinates and sensitive to their signs, leading to the symmetry  between $Q_1$ and $Q_2$ to be broken. In the present study, the symmetry between $Q_1$ and $Q_2$ is broken due to the unequal choice of the parameters corresponding to the two double well potentials (cf. Table~\ref{tab:parameter}) in $V_{0}({\bf Q})$. An additional difference between the models  has to do with the fact that in our case all four minima are degenerate while the Smedarchina model has two pairs of isoenergetic minima.
In any case, our aim here is to study the correspondence between  classical and quantum dynamics of the Hamiltonian model in Eq.~(\ref{eq:hamiltonian}) without necessarily being restricted to a DPT process.
Nevertheless, as seen below, our model shares many of the dynamical features observed in the earlier studies on DPT. 

\begin{figure}
 \centering
 \includegraphics[width=3.5in]{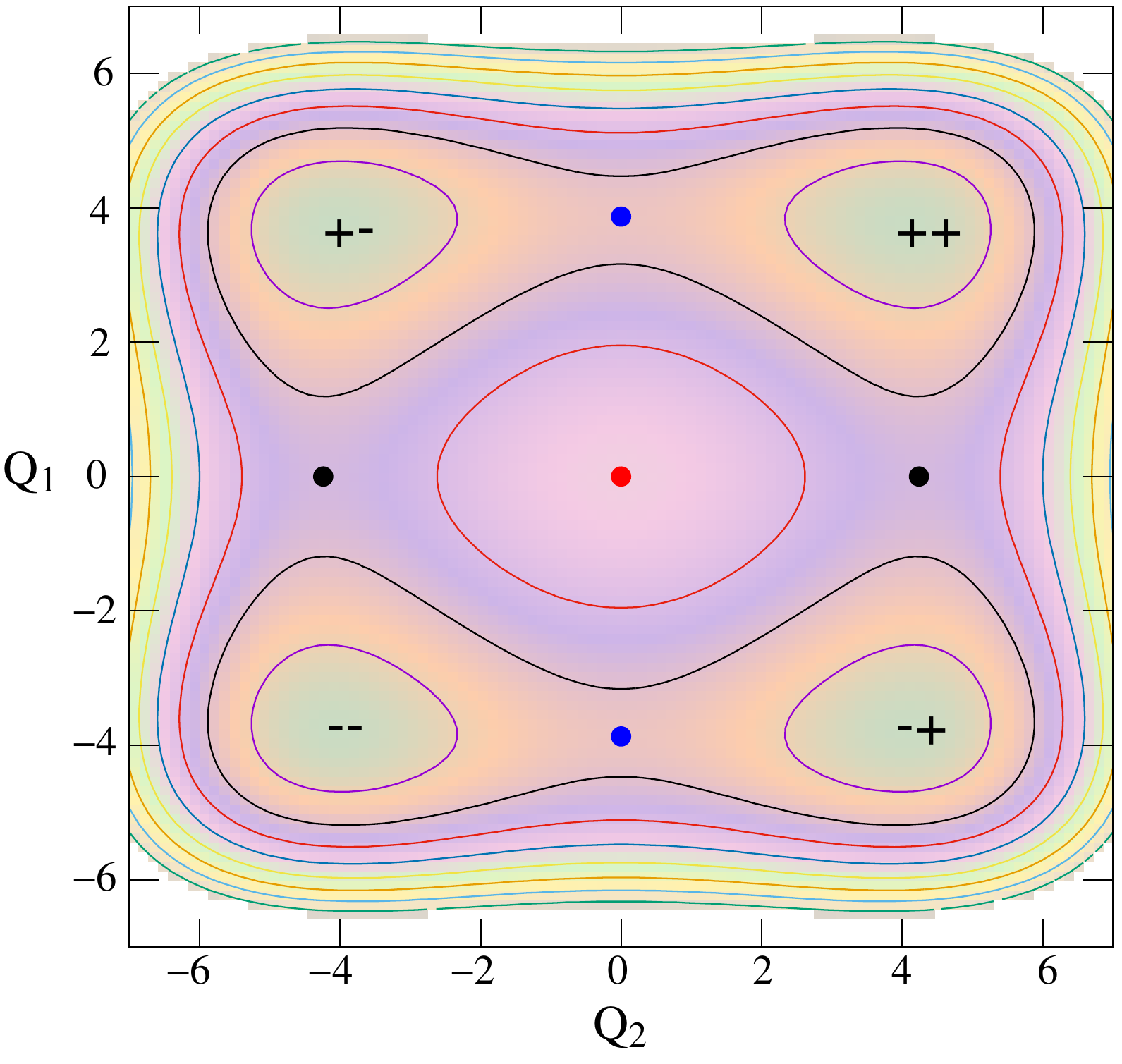}
 \caption{Contour plot of the model potential energy surface. Blue, black, and red points denote the positions of the index-$1$ saddles of lower energy, index-$1$ saddles of higher energy, and the index-$2$ saddle respectively. Symbolic descriptions of the minima refer to the sign of the $(Q_1,Q_2)$ coordinates respectively.}
 \label{fig:pes}
\end{figure}

The numerical values of potential parameters used in this study are listed in Table \ref{tab:parameter}. The values of masses are kept equal to the mass $m_{\rm H}$ of a hydrogen atom. Fig. \ref{fig:pes} represents a contour plot of the potential energy surface (PES). The various critical points of the PES can be summarized as follows:
\begin{enumerate}
    \item Four isoenergetic minima (wells) located at
    \begin{equation}
        (Q_{1},Q_{2}) = \left(\pm \sqrt{\frac{2 a_1 b_2-\gamma a_2}{4 b_1 b_2 - \gamma^{2}}}, \pm \sqrt{\frac{2 a_2 b_1-\gamma a_1}{4 b_1 b_2 - \gamma^{2}}}\right)
    \end{equation}
    Following Collins et al, the minima are denoted as $(--),(+-),(++)$, and $(-+)$ with energy $V_{M}$. In this work we consider the $(--)$ well to be the reactant (R) and the $(++)$ well as the product (P). However, given that the minima are isoenergetic, one can also choose any pair of minima as R and P.
    
    \item The minima are connected by four index-$1$ saddles, of which the two saddles denoted SP$_{1}^{(l)}$ located at 
    \begin{equation}
       (Q_{1},Q_{2}) = \left(\pm \sqrt{\frac{a_1}{2 b_1}},0\right)
    \end{equation}
    have energy $V_{1}^{\dagger}$ lower than the ones with energy $V_{2}^{\dagger}$ located at
    \begin{equation}
       (Q_{1},Q_{2}) = \left(0,\pm \sqrt{\frac{a_2}{2 b_2}}\right)
    \end{equation}
    and denoted by SP$_{1}^{(h)}$.
    
    \item An index-$2$ saddle denoted SP$_2$ with energy $V^{\ddagger} = 0$ at $(Q_1,Q_2)=(0,0)$.
\end{enumerate}

 The position of the various critical points of the PES along with their energies for the specific choice of parameters in Table \ref{tab:parameter} are given in Table \ref{tab:position}. For future reference and for an idea of the timescales involved, we note that an approximate harmonic approximation around the $(--)$ well yields the vibrational time periods of about $0.06$ ps ($\equiv 6 \times 10^{-14}$ s $\equiv 60$ fs) and $0.08$ ps for the $Q_1$ and $Q_2$ modes respectively.

\begin{table}[ht]
\caption{Parameters used in the potential of the system}
\label{tab:parameter}
\setlength{\tabcolsep}{12pt}
\renewcommand{\arraystretch}{1.2}
\begin{tabular}{ll}
\hline
        Parameter &   Value (in au)  \\ 
        \hline
        $a_{1}$ & $3.0\times10^{-3}$ \\ 
        $b_{1}$ & $1.0\times10^{-4}$ \\
        $a_{2}$ & $1.8\times10^{-3}$ \\
        $b_{2}$ & $5.0\times10^{-5}$ \\
        $\gamma $ & $1.0\times10^{-5}$ \\ 
        $m_{\rm H}$ & $1837.151$ \\
        \hline
\end{tabular}
\end{table}

\begin{table}
\caption{Critical points and the corresponding energies (in atomic units (au)) for the model potential}
\label{tab:position}
\setlength{\tabcolsep}{11pt}
\renewcommand{\arraystretch}{1.2}
 \begin{tabular}{l r r r} 
 \hline
        Critical point  & $Q_{1}$ & $Q_{2}$ & Energy  \\ 
        \hline
        {Minima} & $  \pm 3.764$ & $  \pm 4.072$  & $-0.0362$\\
        Index-1 saddle & $ \pm 3.873$ & $  0.0$  & $-0.0225$\\
        Index-1 saddle & $ 0.0$ & $  \pm 4.242$  & $-0.0162$ \\
        Index-2 saddle & $ 0.0$ & $ 0.0$ & $0.0$ \\
        \hline
 \end{tabular}
\end{table}

\section{QUANTUM WAVEPACKET DYNAMICS~\label{sect:quantum_dynamics}}

We begin our dynamical studies by highlighting the key features of the quantum dynamics at energies above the index-$2$ saddle energy. Note that this is motivated by the several quantum dynamical studies that have been done earlier on similar model Hamiltonians in the context of DPT. Our purpose here is to point out the subtleties involved, as emphasized\cite{accardi2010synchronous} by Accardi et al and others\cite{albaredajpcl2015}, in deciphering the reaction mechanism.

\subsection{Choosing the initial state and time-propagation}

The dynamics of the system can be investigated in terms of propagating several types of initial quantum states. However, since our interest lies in exploring the correspondence between the classical and quantum dynamics, we focus here on quantum initial states that are minimum uncertainty wavepackets. Thus, we choose normalized initial states of the form
\begin{equation}
    \Psi({\bf Q},0) = \prod_{j=1,2} N_{j} \exp{\left[-\beta_{j}(Q_{j}-Q_{j_{0}})^2 + \frac{i}{\hbar}P_{j_{0}}(Q_j -Q_{j_{0}})\right]}
    \label{eq:intial_qs}
\end{equation}
corresponding to a wavepacket centered at  (${\bf P}_{0},{\bf Q}_{0}$) with the normalization factor $N_{j} = (2{Re}(\beta_j)/\pi)^{1/4}$. For the wavepacket above the position and momentum uncertainties $\Delta Q_{j}$ and $\Delta P_{j}$ are equal to $1/(2 \sqrt{\beta_{j}})$ and $\hbar \sqrt{\beta_{j}}$ respectively. Thus, $\Delta Q_{j} \Delta P_{j} = \hbar/2$, corresponding to a minimum uncertainty state. In what follows, we set the squeezing parameter $\beta_j = 1.0$, for $j = 1,2$ and vary the $\hbar$.
The mean energy associated with the initial state is evaluated as
\begin{equation}
    \bar{E}  = \langle \Psi|\hat{H}|\Psi \rangle = \sum_{\alpha} |C_{\alpha \Psi}|^{2} E_{\alpha}
\end{equation}
where $C_{\alpha \Psi} \equiv \langle \alpha|\Psi \rangle$ with $|\alpha \rangle$ and $E_{\alpha}$ being the quantum eigenstates and eigenvalues, respectively, of the full Hamiltonian $\hat{H}$.

The time dependent Schr\"{o}dinger equation for a specific initial quantum state $\Psi({\bf Q},0)$
\begin{eqnarray}
    i \hbar \frac{\partial \Psi({\bf Q},t)}{\partial t} &=& \left[\sum_{j=1,2} \left(-\frac{\hbar^{2}}{2m_{\rm H}} \frac{\partial^{2}}{\partial Q_{j}^{2}} + V_{j}(Q_{j})\right) + \gamma Q_{1}^{2} Q_{2}^{2} \right] \Psi({\bf Q},t) \\
    &\equiv& [\hat{K} + \hat{V}] \Psi({\bf Q},t)  
    \label{eq:timedependent}
\end{eqnarray}
is solved numerically using the split operator technique ~\cite{tannor2007introduction,dion2014program}. As this is a well known technique, we provide a brief description of the method.
In Eq.~(\ref{eq:timedependent}), $\hat{K}$ and $\hat{V}$ represent the kinetic and potential operators, respectively.
Given a specific initial quantum state $\Psi({\bf Q},0)$, the quantum state $\Psi({\bf Q},t)$ at time $t$ can be obtained using the unitary time evolution operator $\hat{U}(t)$
\begin{equation}
    \Psi({\bf Q},t)=\hat{U}(t)\Psi({\bf Q},0) \equiv \exp[-i\hat{H}t/\hbar] \Psi({\bf Q},0)
\end{equation}
We can represent $\hat{U}(t)$ as a product of $n$ consecutive time evolution operators over short time intervals $\Delta{t} = t/n$. 
\begin{equation}
   \hat{U}(t) = \underbrace{ e^{-i\hat{H}\Delta{t}/\hbar}e^{-i\hat{H}\Delta{t}/\hbar}\ldots{e^{-i\hat{H}\Delta{t}/\hbar}}}_\text{n times}.
   \label{eq:ntimestep}
\end{equation}
We note that due to the non-commutativity of $\hat{K}$ and $\hat{V}$, we have
\begin{equation}
    {e^{-i[\hat{K} + \hat{V}]\Delta{t}/\hbar}} \approx {e^{-i \hat{K}\Delta{t}/\hbar}}{e^{-i\hat{V}\Delta{t}/\hbar}} + O(\Delta{t}^2)
\end{equation}
The basic idea of the split operator approach is to  introduce a symmetrized product of $\hat{K}$ and $\hat{V}$
\begin{equation}
    {e^{-i\hat{H}\Delta{t}/\hbar}} \approx {e^{-i\hat{V}\Delta{t}/2\hbar}}{e^{-i\hat{K}\Delta{t}/\hbar}}{e^{-i\hat{V}\Delta{t}/2\hbar}} + O(\Delta{t}^3)
    \label{eq:splitapproximation}
\end{equation}
leading to reduced error, and hence faster time evolution.
By using Eq. (\ref{eq:splitapproximation}) in Eq. (\ref{eq:ntimestep}), we obtain 
\begin{equation}
    {e^{-i\hat{H}t/\hbar}} =  {e^{-i\hat{V}\Delta{t}/2\hbar}}\underbrace{{e^{-i\hat{K}\Delta{t}/\hbar}}{e^{-i\hat{V}\Delta{t}/\hbar}}{e^{-i\hat{K}\Delta{t}/\hbar}}{e^{-i\hat{V}\Delta{t}/\hbar}}\ldots{e^{-i\hat{K}\Delta{t}/\hbar}} {e^{-i\hat{V}\Delta{t}/\hbar}}}_\text{n-1 times}{e^{-i\hat{K}\Delta{t}/\hbar}}{e^{-i\hat{V}\Delta{t}/2\hbar}}
\end{equation}
We note that the matrix for $\hat{V}$ is diagonal in the position representation, whereas the matrix for $\hat{K}$ is diagonal in the momentum representation. Consequently, in order to determine the time-evolved state $\Psi({\bf Q},t)$, we use the Fast Fourier transform and its inverse to efficiently propagate the initial state. 
Although it is possible to use more sophisticated splitting algorithms~\cite{blanes2006symplectic,bandrauk1991improved,li2014efficient}, Eq. (\ref{eq:splitapproximation}) is sufficient for the results reported in this study due to the ultrafast nature of the reaction dynamics.  For the calculations reported here, we chose a  $512\times512$ spatial grid. The range for both sets of the spatial  and momentum coordinates is [-8.0,8.0] and [-20.0,20.0].

\subsection{Switch from concerted to non-concerted mechanism}

\begin{figure}[ht]
\includegraphics[width=1.0\textwidth]{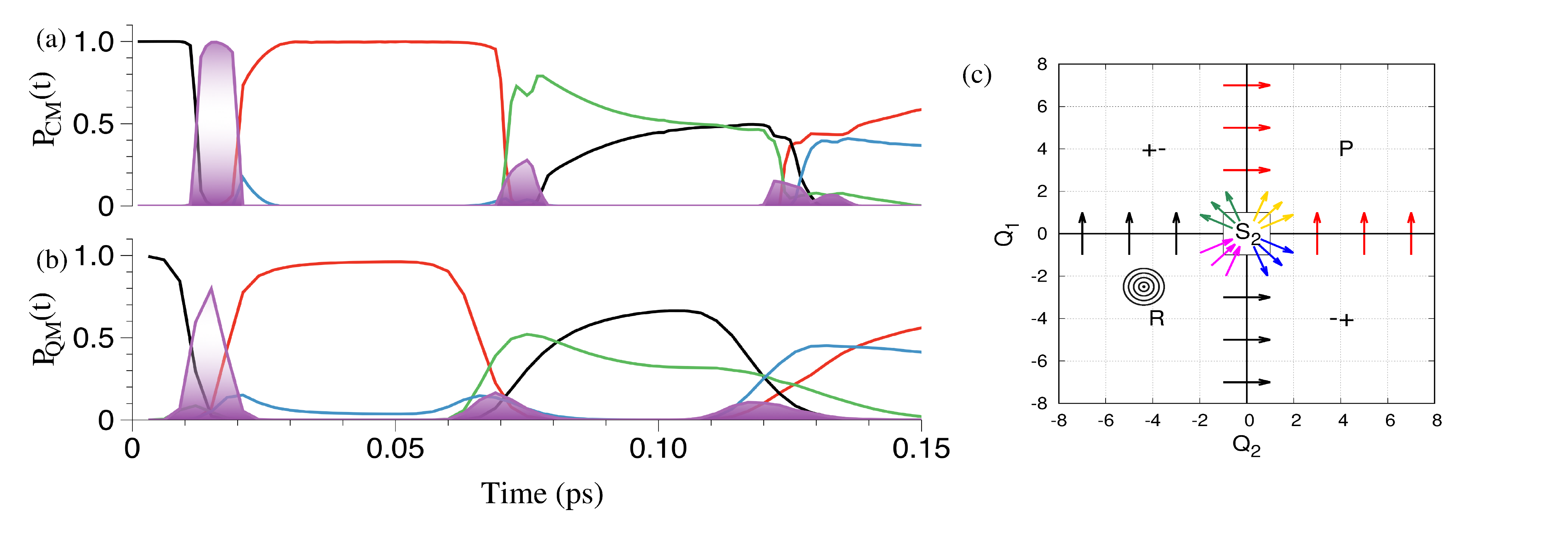}
\caption{(b) Quantum domain probabilities ($\hbar=1.0$) as a function of time for a wavepacket with mean energy of $\approx 0.044$ au. Reactant well $(--)$ (black), product well $(++)$ (red), index-$2$ region, defined by the square region in (c) (magenta filled curve), the $(+-)$ well (green), and the $(-+)$ well (blue). Note the switch in the mechanism from a concerted to a sequential one at $t \sim 70$ fs ($\equiv 0.07$ ps). (a) The corresponding classical probabilities, showing a similar mechanism switch. (c) Definition of the various domains in the configuration space $(Q_1,Q_2)$, adapted\cite{accardi2010synchronous} from an earlier work. The centre of the wavepacket is at $(Q_1,P_1,Q_2,P_2) = (-2.50,9.57,-4.37,14.19)$.}
\label{fig:probswitch}
\end{figure}

Following Accardi et al \cite{accardi2010synchronous}, we define different domains $D$ in the configuration space. As shown in Fig.~\ref{fig:probswitch}(c), we define five domains that include the reactant $(--)$, product $(++)$, ``intermediates" $(+-)$/$(-+)$, and the index-$2$ saddle regions. As noted earlier, the precise geometric definitions of the domains is not particularly important, and the resulting domain probabilities
\begin{equation}
    P_{D}(t) \equiv \int_{D} dQ_1 dQ_2 |\Psi({\bf Q},t)|^{2}
    \label{QM_domainprobs}
\end{equation}
are not significantly different for alternative definitions of $D$. In particular, the mechanistic information is robust to slight changes in the definitions of the domains $D$.

In Fig.~\ref{fig:probswitch}(b), we show the results for the quantum domain probabilities as a function of time for a wavepacket initiated at $(Q_1,Q_2)$ with a mean energy of $E_{wp} \approx 0.044$ au. Two observations are worth noting. First, the reaction is ultrafast and the product regions are populated within a timescale of about $50$ fs, which is of the order of a harmonic vibration period in the reactant well; a similar ultrashort timescale is observed for the average residence time in the intermediate and index-$2$ domains. Second, as noted by Accardi et al\cite{accardi2010synchronous}, the first forward reaction is clearly concerted since the maximum probability is associated with the index-$2$ saddle region. However, at later times ($\sim 150$ fs) the forward reaction is mostly due to a sequential mechanism. In fact, even the first backward reaction from products to reactants is non-concerted. Clearly, Fig.~\ref{fig:probswitch}(b) indicates that the role of the index-$2$ saddle for the overall reaction is gradually reduced with time. 

Given the fact that the  mean energy of initial wavepacket is much above the index-$2$ saddle energy, the switching of the reaction mechanism from a concerted to a sequential one is perhaps puzzling. We note here that due to the ultrashort residence times in the $(+-)/(-+)$ and index-$2$ domains, it has been argued that the mechanism can be considered as effectively synchronous. Nevertheless, rationalizing the observations in Fig.~\ref{fig:probswitch}(b) is crucial for two main reasons. Firstly, such an insight may offer interesting clues for implementing rational control strategies that exploit the presence of higher index saddles. Secondly, the dynamical switching of the mechanism might pose a challenge for TST-based rate computations in terms of fluxes across appropriately defined locally recrossing free dividing surfaces.

\begin{figure}[ht]
\includegraphics[width=1.0\textwidth]{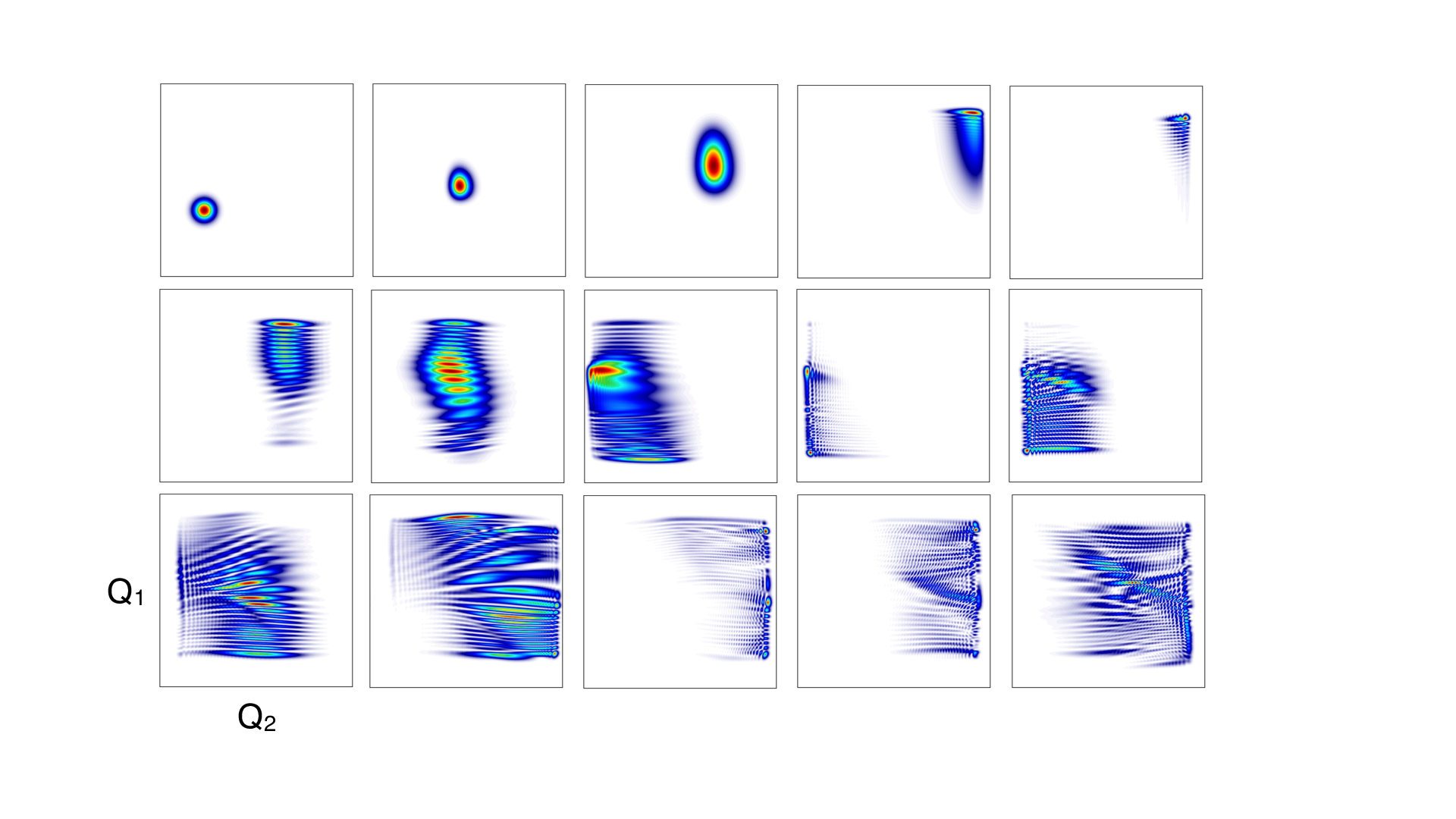}
\caption{Configuration space snapshots of the time evolving wavepacket corresponding to Fig.~\ref{fig:probswitch}(b). The top column left panel corresponds to time $t=0$ ps and the right column bottom panel corresponds to $t=0.168$ ps. The rest of the panels are shown at intervals of $0.012$ ps.}
\label{fig:wavepacket_snap_chaotic}
\end{figure}

In order to understand the mechanism switch, in Fig.~\ref{fig:wavepacket_snap_chaotic} we show the time evolution of the wavepacket which results in the domain probabilities shown in Fig.~\ref{fig:probswitch}(b). Starting from the initial time the evolution is shown in time steps of $0.012$ ps until a final time of $0.168$ ps. Note that this time range, as evident from Fig.~\ref{fig:probswitch}(b), covers the first forward and backward reaction and the second forward reaction.  The various aspects of the time evolution, from the initial flux across the index-$2$ region to the subsequent wavepacket dispersion and relief reflections leading to the domain densities in Fig.~\ref{fig:probswitch}(b), are in accordance with the earlier studies. Consequently, as analyzed and argued earlier in detail\cite{accardi2010synchronous}, the switch from a concerted to the stepwise mechanism is due to the dispersion and the subsequent relief reflections of the wavepacket from the steep repulsive wall of the PES in the product domain into the $(-+)/(+-)$ domains.

\begin{figure}[ht]
\includegraphics[width=1.0\textwidth]{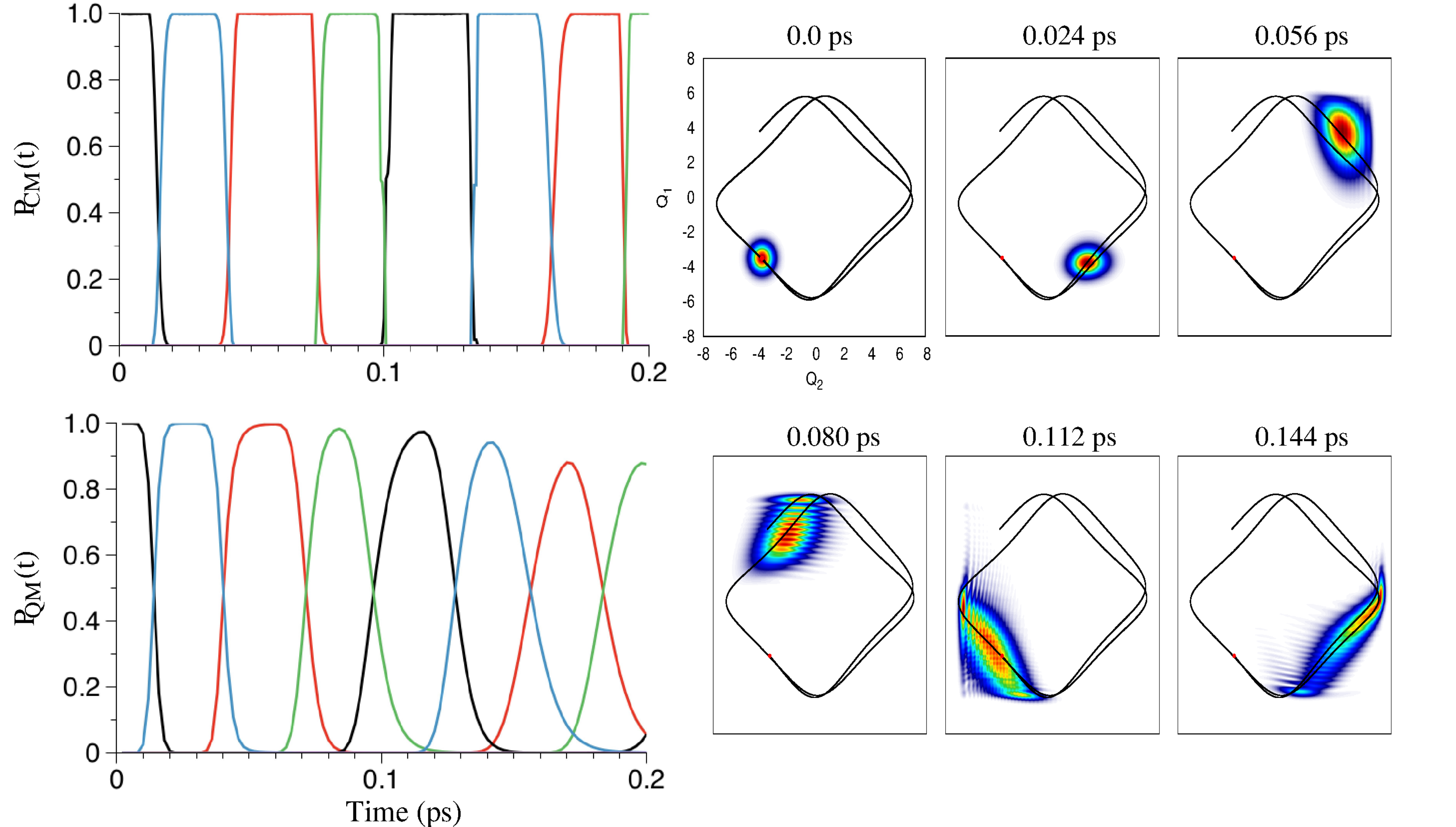}
\caption{Initial quantum wavepacket with mean energy $\bar{E} \approx 0.045$ au exhibiting Murrell-Laidler type dynamics. The associated domain probabilities (a) classical and (b) quantum are shown. Configuration space snapshots of the wavepacket at indicated times are also shown. For reference, the classical trajectory corresponding to the center of the wavepacket is shown as a bold black line. The centre of the wavepacket is at $(Q_1,P_1,Q_2,P_2) = (-3.50,-12.07,-3.77,12.97)$.}
\label{fig:DML_probs_wpsnaps}
\end{figure}

\subsection{Is the mechanism switch a quantum effect?}

Given the results in Fig.~\ref{fig:probswitch}(b) and the wavepacket dynamics shown in Fig.~\ref{fig:wavepacket_snap_chaotic} along with the rationale provided above, it is reasonable to expect that the switching of the mechanism is of quantum origins. However, confirming such expectations requires investigating the corresponding classical dynamics. Thus, we compute the classical analog of the domain probabilities in Eq.~(\ref{QM_domainprobs}) as
\begin{equation}
    P_{D}^{cm}(t) = \int_{D} d{\bf Q} \int d{\bf P} \, \rho_{cm}({\bf Q},{\bf P};t)
    \label{CM_domainprobs}
\end{equation}
with the classical phase space density computed based on the formula~\cite{helmkamp1994structures}
\begin{equation}
    \rho_{cm}({\bf Q},{\bf P};t) = \int d{\bf Q}' d{\bf P}' \delta[{\bf Q}-{\bf Q}_{t}({\bf Q}',{\bf P}')] \delta[{\bf P}-{\bf P}_{t}({\bf Q}',{\bf P}')] \rho_{cm}({\bf Q}',{\bf P}';0)
\end{equation}
In the above formal solution of the Liouville equation, $({\bf Q}_{t},{\bf P}_{t})$ is the classical trajectory with the initial condition $({\bf Q}',{\bf P}')$. The classical density at $t=0$ is chosen as Gaussians in the phase space with position and momentum widths consistent with the initial quantum wavepacket density $|\Psi({\bf Q},0)|^{2}$. 
In order to compute $P_{D}^{cm}(t)$ we initiate an ensemble of $20000$ initial conditions sampled according to the initial density $\rho_{cm}({\bf Q}',{\bf P}';0)$ and integrate their equation of motion forward in time for $300$ fs ($0.3$ ps) using a fourth order Runge-Kutta algorithm.

The resulting classical domain probabilities, shown in Fig.~\ref{fig:probswitch}(a), clearly indicate that the mechanism switch occurs classically as well. Moreover, apart from the expected quantitative differences, the classical and quantum results agree with regards to the timescales and the overall qualitative dynamics. This is perhaps not surprising, since the dynamics of minimum uncertainty wavepackets on short timescales are generally expected to exhibit excellent classical-quantum correspondence. Nevertheless, it behooves us to identify the origins of the mechanism switch from a classical phase space perspective. 

Note that several other wavepackets, with mean energy greater than that of the index-$2$ saddle, exhibit similar switching dynamics. However, it is crucial to point out that there do exist initial wavepackets that undergo very different dynamics. In particular, they completely ignore the index-$2$ region, instead utilizing the $(-+)/(+-)$ domains i.e., a stepwise mechanism. Such examples can be labeled as ``dynamical" Murrell-Laidler (DML) cases - dynamical, since the trajectories are not on the IRC, and Murrell-Laidler since they access only the domains corresponding to the index-$1$ saddles. In Fig.~\ref{fig:DML_probs_wpsnaps}, we show one such example for DML for an initial state with a mean energy $\bar{E} \approx 0.045$ au. Again, the classical and quantum domain probabilities are in good qualitative agreement and the avoidance of the index-$2$ region is clear. The snapshots of the wavepacket shown in Fig.~\ref{fig:DML_probs_wpsnaps} indicate that the wavepacket center essentially follows the classical trajectory. 

We now turn our attention to a detailed investigation of the manifolds in the classical phase space that are responsible for the observed reaction dynamics.

\section{CLASSICAL DYNAMICS: DYNAMICAL MURRELL-LAIDLER AND THE COMPETITION BETWEEN STEPWISE AND CONCERTED MECHANISMS}

\subsection{A survey of the classical phase space}

The model Hamiltonian in Eq.~(\ref{eq:hamiltonian}) has two degrees of freedom and hence one can use the Poincar\'{e} surface of section to visualize the global phase space structures.
However, due to the multi-well potential, a given sectioning condition can only reveal part of the dynamical information. For example, the $Q_1=0$ section will show the trajectories that isomerize between the $(--)$ and $(+-)$ wells but not the transitions between the $(--)$ and $(-+)$ wells. Clearly, any sectioning condition will only reveal two among the four wells. Therefore, some care is required in interpreting the surface of sections.

\begin{figure}[ht]
\includegraphics[width=1.0\textwidth]{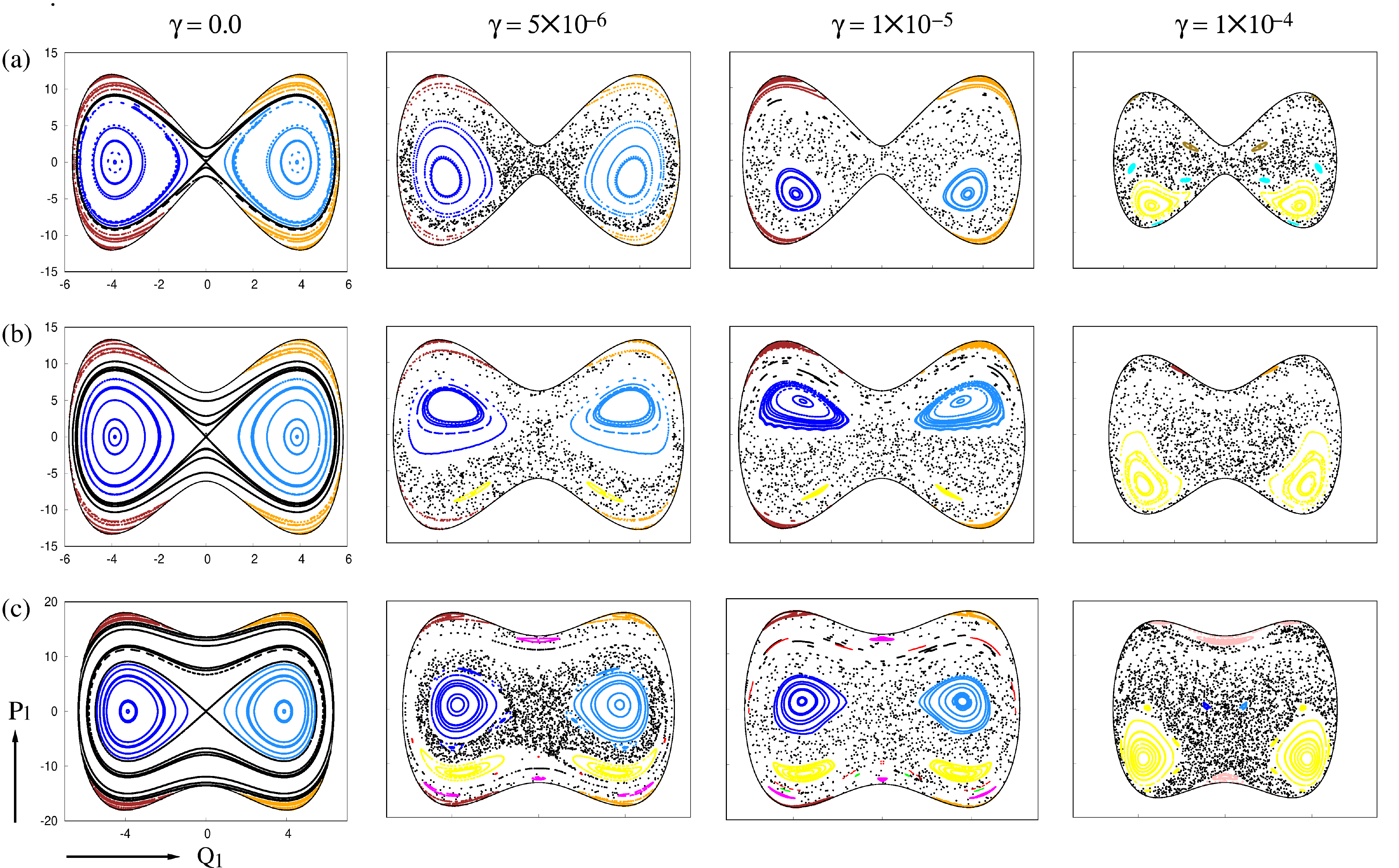}
\caption{Poincar\'{e} surface of sections along $Q_{1} = Q_{2}$ for various coupling values (indicated) at a total energy of (a) $0.001$ au (top row) (b) $0.01$ au (middle row) (c) $0.05$ au (bottom row). The outer black line in the plots is the boundary of energetically accessible energy surface, that is, the zero velocity curve or \emph{Hills region}. In addition, we note that symmetric island regions with different color mean that they are classically disconnected.}
\label{fig:psos}
\end{figure}

\begin{figure}[ht]
\includegraphics[width=0.75\textwidth]{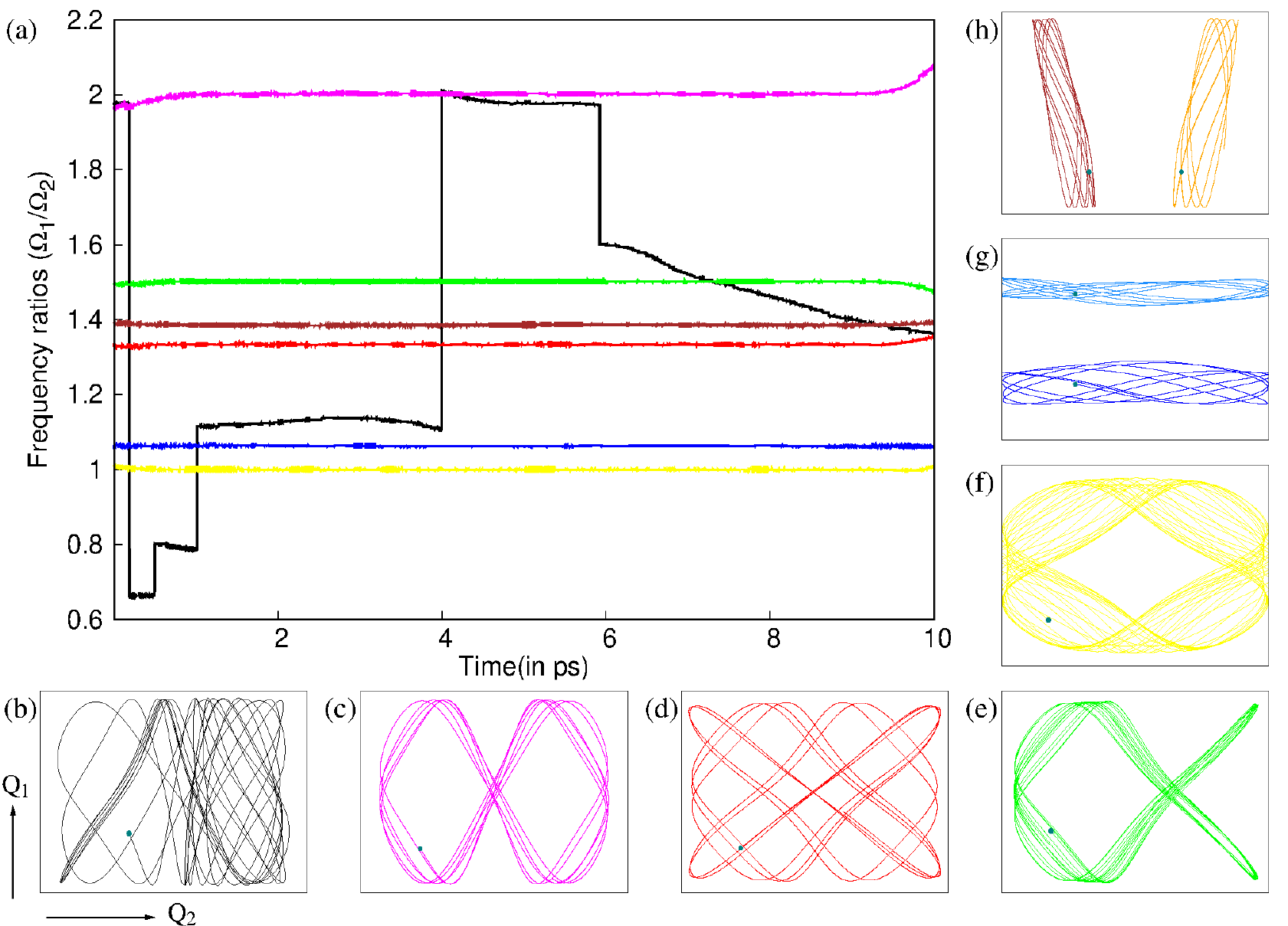}
\caption{(a) Time-frequency analysis for the trajectories present at total energy of $0.05$ au and coupling strength $\gamma = 1 \times 10^{-5}$. Configuration space representation of (b)-(e) concerted and (f)-(h) non-concerted trajectories. Note that the color code is the same as indicated in the corresponding surface of section shown in Fig.~\ref{fig:psos} and the teal colored circle represents the initial state of the trajectory }
\label{fig:config_plots}
\end{figure}

Since in this work we are interested in the isomerization between the $(--)$ (reactant) and the $(++)$ (product) wells, we choose the $Q_{1} = Q_{2}$ sectioning condition. In Fig.~\ref{fig:psos}, we show the resulting Poincar{\'e} sections for three values of the total energy that are above that of the index-$2$ saddle and for coupling strengths ranging from $\gamma=0$ to $\gamma = 1.0 \times 10^{-4}$. We show example configuration space representation of the different dynamics in Fig.~\ref{fig:config_plots} and note several observations at this stage. Firstly, for a fixed energy, increasing the coupling leads to a mixed regular-chaotic phase space. Secondly, one can observe several classes of regular trajectories, particularly  at the higher energies. Clearly, the phase space is far from being completely chaotic even for the highest energy and coupling strength considered.  In turn this observation, apart from illustrating the danger of naively assuming ergodicity at sufficiently high energies and couplings,  implies that whether the mechanism is concerted or not depends crucially on the nature of the initially prepared state. As evident from the configuration space representation of the various trajectories in Fig.~\ref{fig:config_plots}, even at $E = 0.05$ au there is a coexistence of both concerted and non-concerted dynamics.

Note that, amongst the regular trajectories in Fig.~\ref{fig:psos}, several nonlinear resonances can be observed. Such nonlinear resonances correspond to facile energy exchange between the two modes. A particularly interesting example is the  prominent resonance shown in yellow in Fig.~\ref{fig:psos} whose configuration space representation (cf. Fig.~\ref{fig:config_plots}(f)) shows that they correspond to the DML type. In particular, the trajectory shown in Fig.~\ref{fig:DML_probs_wpsnaps} and the associated wavepacket evolution establish that purely non-concerted mechanism can manifest well above the index-$2$ saddle energy, depending on the initial state preparation. It is important to note that the DML mechanism occurs due to the coupling $\gamma \neq 0$. This fact emphasizes the crucial role of IVR in the reaction mechanism.

We note that the various nonlinear resonances that may manifest in the phase space can be predicted based on the zeroth-order Hamiltonian. This is possible since the exact zeroth-order nonlinear frequencies can be computed for a quartic oscillator. However, as is well known, different sets of action-angle variables and frequencies are associated with energies below and above the index-$1$ saddles. Thus, as is relevant to the DML example, focusing on the case wherein the unperturbed energies for both the modes are greater than the respective index-$1$ saddle energies, the ratio of the exact unperturbed frequencies is given by the expression
\begin{equation}
    \frac{\Omega_{1}}{\Omega_{2}} = \left[\frac{a_{1}^{2} + 4 b_{1} E_{1}}{a_{2}^{2} + 4 b_{2} E_{2}}\right]^{1/4} \frac{\mathbb{K}(k_{1})}{\mathbb{K}(k_{2})}.
\end{equation}
In the above equation, $E_1$ and $E_2$ correspond to the zeroth-order energies in the two modes. The $\mathbb{K}(k)$ are complete elliptic integrals of the first kind with
\begin{equation}
    k_{j}^{2}(E_{j}) = \frac{1}{2} \left[1 + \frac{a_j}{\sqrt{a_{j}^{2} + 4 b_{j} E_{j}}}\right]
\end{equation}
As usual, for $\Omega_1/\Omega_2 = r/s$ with integers $(r,s)$  one predicts the specific nonlinear resonances, visible in the surface of section as resonance islands. Note that the analytic solution $q_{j} \propto cn(2\mathbb{K}(k_{j}) \theta_{j}/\pi,k_{j})$, with $\dot{\theta}_{j} \equiv \Omega_{j}$, allows one to determine the  various possible nonlinear resonances. Indeed, using the standard Fourier representation of the Jacobi elliptic function it is possible to show that the primary resonances are of the form $\Omega_{1}:\Omega_{2} = 2r:2s$. Consequently, the lowest order term is a $2:2$ resonance which corresponds to the DML dynamics. Similar results can be derived for $E_{1}$ and $E_{2}$ being below the index-$2$ saddle energies, only now all resonances of the form $\Omega_{1}:\Omega_{2} = r':s'$ are possible.

For the model Hamiltonian of interest further analysis in terms of the zeroth-order action-angle variables is not straightforward since the zeroth-order Hamiltonian cannot be written down explicitly in terms of the action variables. However, numerically, the technique of time-frequency analysis is a powerful approach to determine the various frequency lockings. Moreover, the time-frequency analysis can be applied rather generally to the fully coupled system in order to analyze the non-integrable dynamics. Here we use the continuous wavelet transform approach~\cite{vela2001time} which has been employed earlier in several studies of IVR and reaction dynamics\cite{ksacp2013}. Using this approach, in Fig.~\ref{fig:config_plots} we show the frequency ratios for the various classes of dynamics that are possible in the model system. As expected, chaotic trajectories have frequencies that vary considerably with time whereas the regular trajectories have constant frequencies. Nevertheless, an important aspect to note from the time-frequency analysis is that the chaotic trajectories exhibit stickiness on fairly long timescales. Interestingly, from the lone example of chaotic case shown in Fig.~\ref{fig:config_plots}, the stickiness can occur near concerted as well as non-concerted phase space features.

\subsection{Unstable periodic orbits and invariant manifolds that regulate the reaction dynamics}

In this section, we present the phase space structure governing the mechanism of concerted and sequential isomerization. As discussed in Sect.~\ref{sect:quantum_dynamics}, the switch from concerted to sequential is accessible only when the energy is above the energy of the index-2 saddle. We have shown that using quantum wavepacket and classical density calculations and here we will present the underlying phase space structures for total energy, $E > E_{\rm S_2}$, where $E_{\rm S_2}$ denotes the energy of the index-2 saddle.

\begin{figure}
    \centering
    \includegraphics[width=0.48\textwidth]{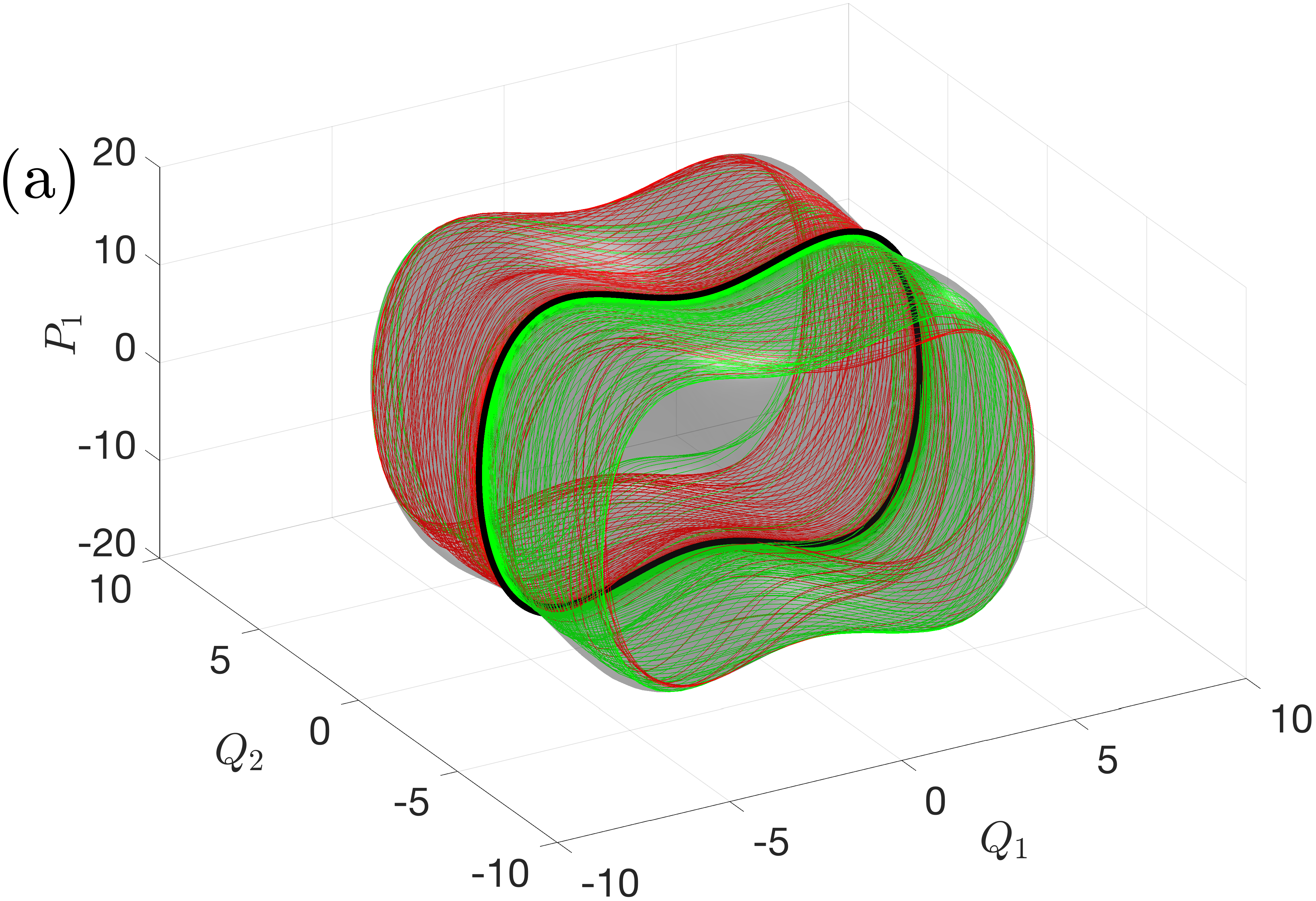}
    \includegraphics[width=0.48\textwidth]{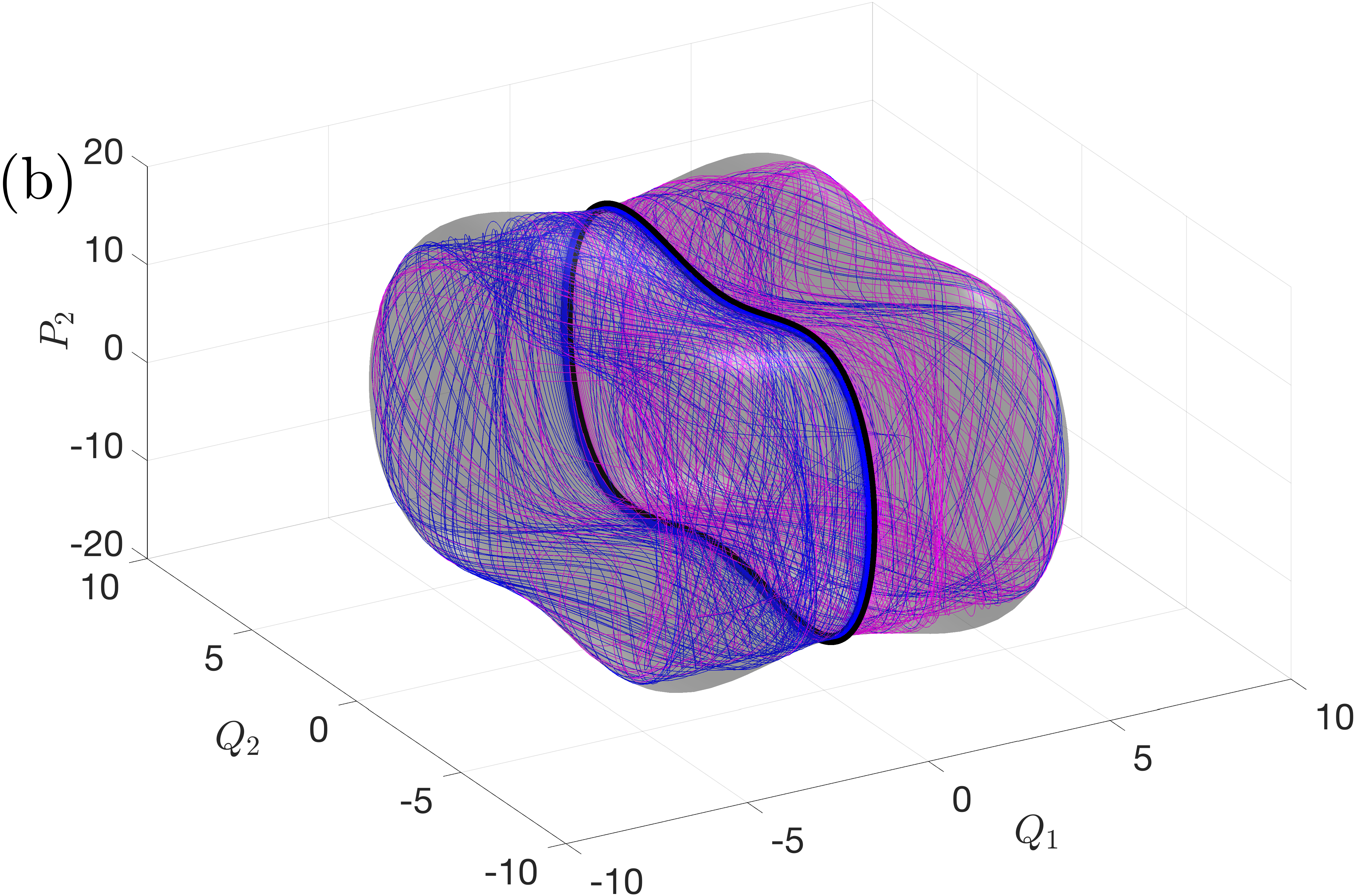}
    \caption{Phase space structures underlying the reaction mechanism switch at energies above the energy of the index-2 saddle. Shown here are unstable periodic orbits, $\Gamma_{13}$ and $\Gamma_{24}$, and its associated invariant manifolds in (a) and (b), respectively, and which are co-dimension 1 in the energy surface. The pair of cylindrical manifolds form the skeleton of the reaction mechanism switch. The manifolds have been globalized for $t = 0.5$ ps and at $E = 0.045$ au which corresponds to the mean energy of the wavepacket used in quantum dynamics calculations.}
    \label{fig:E45e-3_gamma1e-5_time15e-2_upos_tubes}
\end{figure}

As the energy is increased above the energy of the index-2 saddle, the unstable periodic orbits associated with the index-1 saddles with $Q_2 = 0$ coordinates (at energy $-0.0225$ au) coalesce to form one unstable periodic orbit, $\Gamma_{13}$. Similarly, another unstable periodic orbit, $\Gamma_{24}$, is obtained due to the coalescence of index-1 saddles with $Q_1 = 0$ (at energy $-0.0162$ au) coordinates. The unstable periodic orbits are obtained using a differential correction of guess initial conditions and numerical continuation to obtain the orbit at the desired energy. The unstable periodic orbits have associated invariant manifolds of geometry $\mathbb{S} \times \mathbb{R}^1$ (cylindrical geometry or tubes), as shown in Fig.~\ref{fig:E45e-3_gamma1e-5_time15e-2_upos_tubes}. These are computed using globalization of the initial conditions displaced along the corresponding eigenvectors, which is used to generate the intersections with the surface of section (Eqn.~\ref{eqn:cartU1}) and the resulting homoclinic tangle. To justify these phase space structures as underlying the reaction mechanism switch, we present the results for $E = 0.045$ au and $\gamma = 1 \times 10^{-5}$ au to compare with the wavepacket calculations. We note that our discussion holds as long as the stability type of the periodic orbits does not change with total energy and coupling strength. A detailed analysis of the changes in the stability and related manifestation of the reaction mechanism is beyond the scope of this article and will be the focus of future work. The methods used for computing the phase space structures are described in the supplemental material and we show the homoclinic tangles for sample values of the total energy and coupling strength. Let us define the surface of section: 
\begin{equation}
	U_1^{\pm} = \{ (Q_1, P_1) \; | \; Q_2 - Q_1 = 0, \; -{\rm sign}(P_2 - P_1) = \pm 1  \} \label{eqn:cartU1}
\end{equation}
where the $\pm$ indicates direction of crossing of the surface and in this article, we use the $+$ direction. We note that a detailed analysis of phase space transport in this system requires combining the two diagonal sections with appropriate crossing direction. 

\begin{figure}
    \centering
    \includegraphics[width=1.0\textwidth]{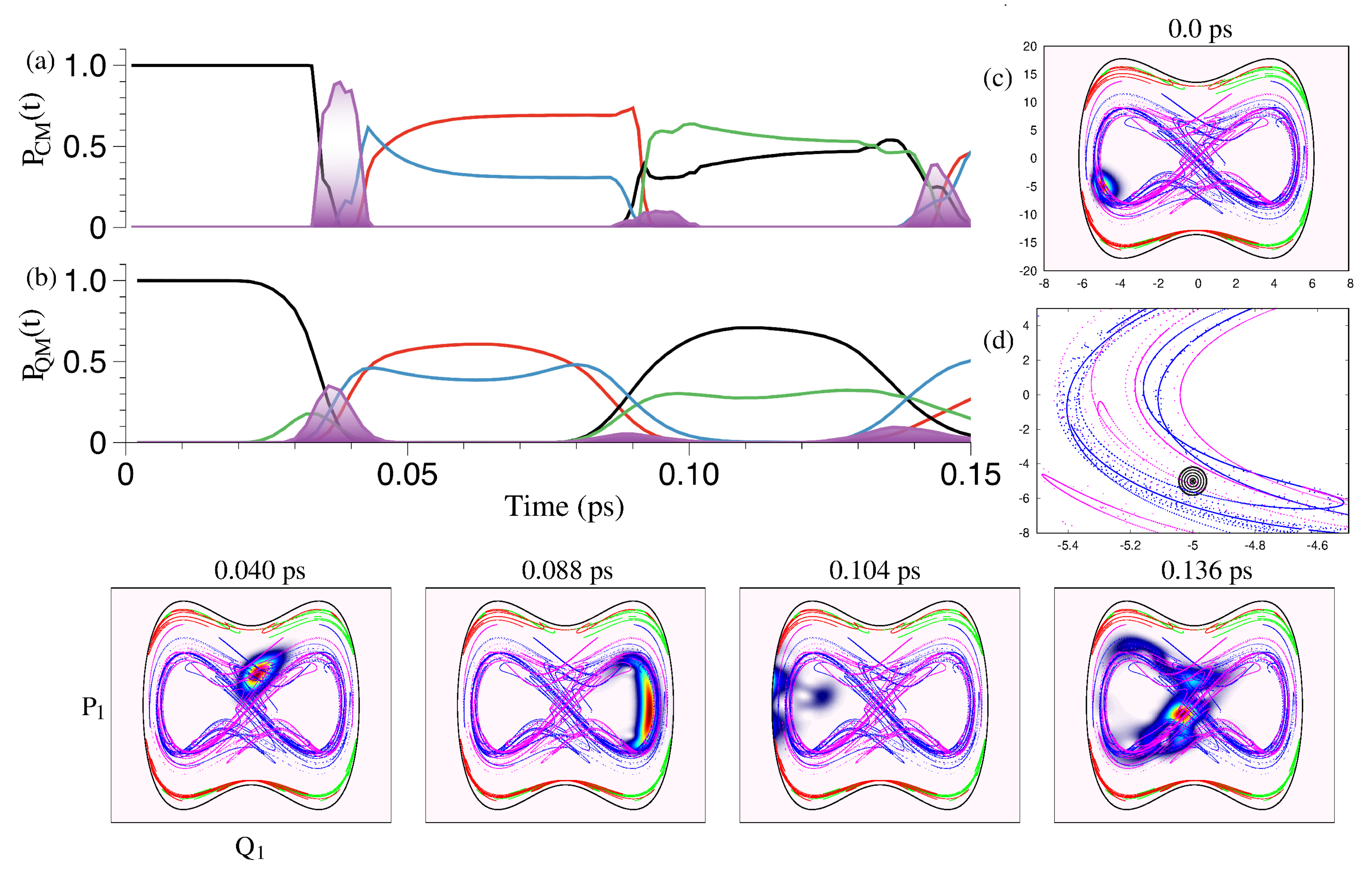}
    \caption{(a) Classical and (b) quantum ($\hbar = 1.0$) domain probabilities  for an initial condition chosen on the homoclinic tangle. (c) Location of the Husimi distribution at $t=0$ superposed on the stable (blue) and unstable (magenta) manifolds of the unstable periodic orbit $\Gamma_{24}$. (d) A zoomed version of (c) indicating the precise location of the initial wavepacket. Four different time snapshots of the evolving Husimi distributions are shown. The centre of the wavepacket is at $(Q_1,P_1,Q_2,P_2) = (-5.00,-5.00,-5.00,-14.62)$.}
    \label{fig:tangle_ic1_combined}
\end{figure}

Due to the geometry of the unstable periodic orbits (UPOs) and the location of the surface of section, $\Gamma_{24}$ appears as a point ($Q_1 = 0, P_1 = 0$) and $\Gamma_{13}$ appears as two points (located near the energy boundary at $Q_1 = 0$) on the surface of section.  In addition, the pair of invariant manifolds associated with $\Gamma_{13}$ and $\Gamma_{24}$ partition the energy surface into dynamically distinct regions. However, due to their co-existence at the same energy, their combined role in phase space transport of classical trajectories is to be expected\cite{wiggins_role_2016}. An easier way to explain and visualize their interplay is via the intersection of the manifolds with the surface of section. The stable and unstable invariant manifolds of these UPOs form the homoclinic tangle as shown in Fig.~\ref{fig:tangle_ic1_combined}(c) and~\ref{fig:tangle_ic2_combined}(c).

\subsection{Modulating the role of the index-$2$ saddle}

The above construction of the invariant manifolds, in light of the good classical-quantum correspondence observed for the domain probabilities in Fig.~\ref{fig:probswitch}, raises the intriguing possibility of modulating the extent to which the index-$2$ saddle influences the reaction dynamics. We note that the stable and unstable invariant manifolds of the UPOs form the homoclinic tangle, leading to large scale chaos on sufficiently long timescales. However, the ultrashort  dynamics implies early time structure in the homoclinic tangle and, related to the early work on reactive islands~\cite{marston1989reactive}, specific initial states may exhibit widely different mechanisms.  We note, however, that there is a distinct possibility that the quantum wavepacket  tunneling can compromise the classical ``impenetrable" barriers. Apart from the domain probabilities, we perform a more detailed comparison between the classical and quantum dynamics in terms of computing the Husimi distribution~\cite{husimi1940some}, which is a phase space distribution associated with the quantum state $\Psi({\bf Q},t)$ 
\begin{equation}
    \rho_{H}({\bf P}_0,{\bf Q}_0;t) = \frac{1}{(2\pi\hbar)^2} |\langle \phi|\Psi(t) \rangle|^2
\end{equation}
 where $\langle {\bf P}_0,{\bf Q}_0|\phi \rangle$ is a minimum uncertainty state of the form in Eq.(~\ref{eq:intial_qs}) centered at $({\bf P}_0,{\bf Q}_0)$. Here we calculate the projection of the Husimi distribution on to the $Q_1=Q_2$ plane, in order to be consistent with the classical surface of section.

\begin{figure}
    \centering
    \includegraphics[width=1.0\textwidth]{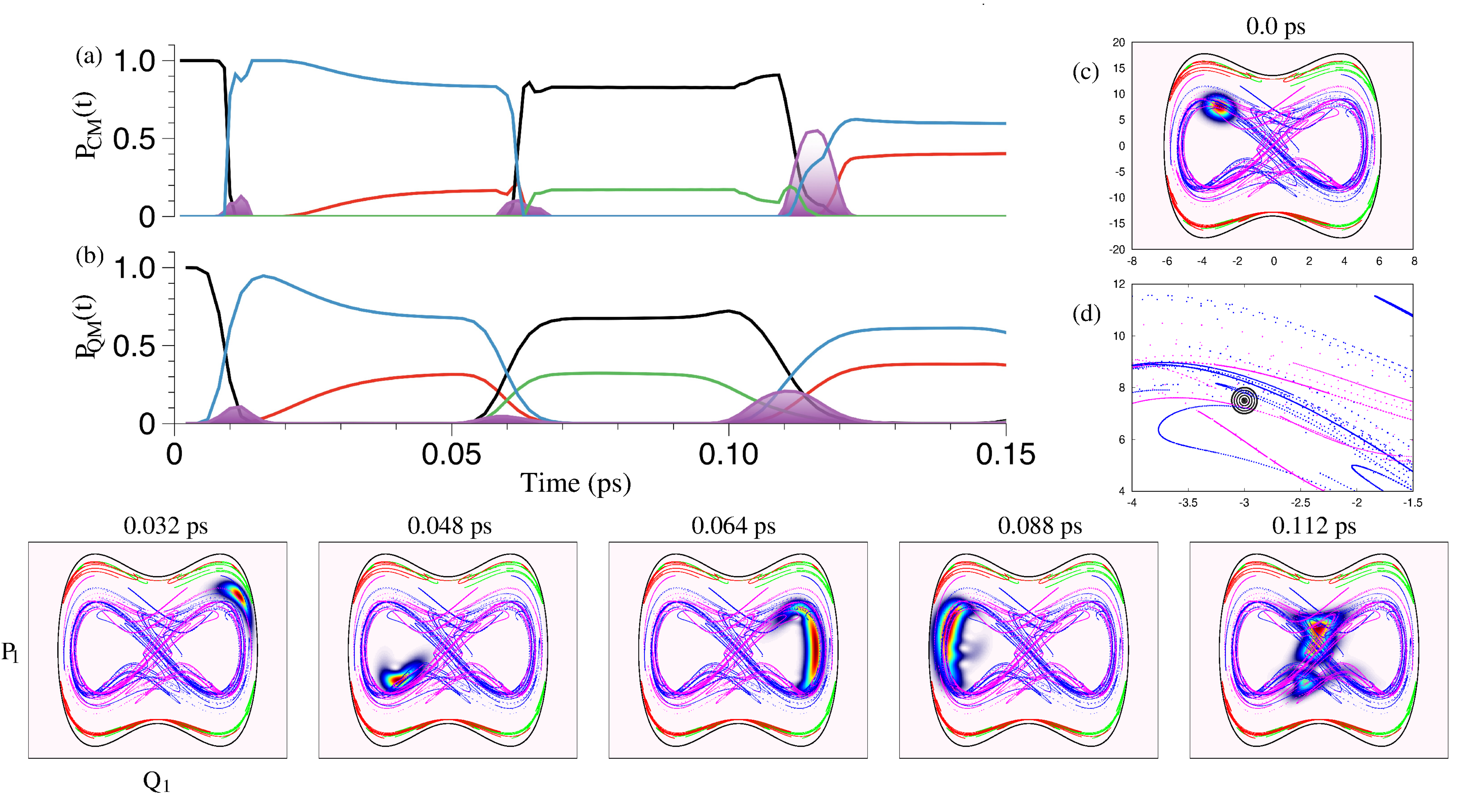}
    \caption{(a) and (b) Classical and the quantum ($\hbar = 1.0$) domain probabilities respectively for another initial condition chosen on the homoclinic tangle. (c) Location of the Husimi distribution at $t=0$ superposed on the stable (blue) and unstable (magenta) manifolds of the unstable periodic orbit $\Gamma_{24}$. (d) A zoomed version of (c) indicating the precise location of the initial wavepacket. Four different time snapshots of the evolving Husimi distributions are shown. The centre of the wavepacket is at $(Q_1,P_1,Q_2,P_2) = (-3.00,7.50,-3.00,14.84)$.}
    \label{fig:tangle_ic2_combined}
\end{figure}

To consolidate our claim, we visualize the manifold intersections on the surface of section along with the evolution of the Husimi distribution over the same time interval for two different initial conditions. In  Fig.~\ref{fig:tangle_ic1_combined} and~\ref{fig:tangle_ic2_combined}(c) we show the location of the center of the quantum wavepacket with respect to the homoclinic tangle formed by the intersection of the stable and unstable invariant manifolds associated with the UPOs. A closer look at the location of the initial wavepackets  can be seen in Fig.~\ref{fig:tangle_ic1_combined} and~\ref{fig:tangle_ic2_combined}(d). It is interesting to note that despite both the initial conditions being in the homoclinic tangle and at the same energy of $E = 0.045$ au, the reaction mechanisms are quite different. Moreover both the initial conditions also exhibit mechanisms  distinct from the example shown in Fig.~\ref{fig:probswitch} in terms of the extent of involvement of the index-$2$ saddle.  Specifically, the initial wavepacket in Fig.~\ref{fig:tangle_ic2_combined} shows that the index-$2$ saddle starts to play a role at a much later time $t \sim 110$ fs. In this case the early time mechanism is dominantly sequential or stepwise $(--) \rightarrow (-+) \rightarrow (++)$. In contrast, the dynamics in Fig.~\ref{fig:tangle_ic1_combined} clearly involves the index-$2$ saddle at a much earlier time. However, there is an important difference between the mechanism for the initial wavepacket in Fig.~\ref{fig:tangle_ic1_combined}(b) and the one in Fig.~\ref{fig:probswitch}(b). Although both do involve the index-$2$ saddle dominantly at early times, the former case utilizes the concerted dynamics for the $(+-) \rightarrow (-+)$ transfer and then sequentially to the product whereas in the latter case the index-$2$ saddle's role is  predominantly to lead to the $(--) \rightarrow (++)$ population transfer. 

The snapshots of the Husimi distributions shown in the bottom panels of Fig.~\ref{fig:tangle_ic1_combined} and~\ref{fig:tangle_ic2_combined}  indicates that the evolution is mainly along the ``track'' formed by the homoclinic tangle of the $\Gamma_{13}$ UPO. This indicates, apart from justifying the close agreement between the classical and quantum domain probabilities, a fairly detailed classical-quantum correspondence when it comes to the transport in the phase space. In other words, much of the observed mechanisms are justifiable from a purely classical dynamical perspective. This observation further bolsters the claim by Accardi et al\cite{accardi2010synchronous} that the wavepacket mechanism switching is essentially a quantized form of the mechanisms observed by Black et al\cite{black2012dynamics}.
However, it is worth pointing out that despite this reasonably good classical-quantum correspondence,  one can clearly see key differences between the classical and quantum dynamics. Firstly,  as gleaned from the various domain probabilities shown, the role of the index-$2$ saddle is more dominant classically when compared to the quantum dynamics. Secondly, signatures of quantum tunneling can be observed both in the domain probabilities and in the time evolution of the Husimi distributions.  In case of the domain probabilities the tunneling effects are, for instance, seen at the various transition times - the classical probabilities exhibit a near ``step" function drop when compared to the more smoother quantum variations. In terms of the Husimi a clear indication of quantum tunneling can be seen from the snapshot in Fig.~\ref{fig:tangle_ic1_combined} at $t \sim 0.104$ ps wherein lower density patches can be observed in the reactant well. Note that around this time the classical and the quantum domain probabilities are also not in a good correspondence.

\section{CONCLUSIONS AND OUTLOOK}

Our dynamical studies on a  two degrees of freedom system model for double proton transfer in molecular systems has clearly established a good classical-quantum correspondence with regards to the  mechanism of the reaction. More importantly, our studies corroborate the expectation that the dynamical complexity precludes a clean separation of the mechanism into purely sequential or concerted one. We also show that the phase space of the system, despite strong couplings, continues to exhibit mixed regular-chaotic dynamics. As a consequence, the mechanism is expected to be sensitive to the specific molecule and the prepared initial state.

As we noted in the discussion in the last subsection, the detailed classical-quantum correspondence and the ultrafast nature of the dynamics may allow for modulating the influence of the index-$2$ saddle to the overall reaction mechanism. Although such an expectation is borne out by Figs.~\ref{fig:tangle_ic1_combined} and \ref{fig:tangle_ic2_combined}, extracting the detailed phase space mechanisms warrant further studies. In particular, the differences in the reaction mechanism via crossing the index-2 saddle between the two initial conditions sampled from the homoclinic tangle are expected to be due to the result of the underlying structure of the lobes~\cite{wiggins1990geometry}. We suspect that this, and the reaction mechanism switch observed in Fig.~\ref{fig:probswitch}, can be investigated by using lobe dynamics on the section~\eqref{eqn:cartU1} along with the section passing through the intermediate wells. In this context a crucial question is whether experiments can prepare such exquisitely tailored wavepackets in complex systems. An equally relevant question is whether the predictions of the lobe dynamics are robust in the presence of noise and or dissipation. Further work is required to address both of the above issues. 

Finally, there are indications that coupling of at least another mode to the two degree of freedom DPT model is crucial\cite{walewski2010car}. The results of the current study suggest that the classical-quantum correspondence should hold in the higher degree of freedom system as well. However, identifying the invariant manifolds and transport barriers in phase space will require techniques such as that of Lagrangian descriptors on suitably chosen sections~\cite{mendoza2010hidden,patra2018detecting,mancho2013lagrangian,garcia2020exploring} and the recently introduced asymptotic trajectory indicator method\cite{nagahata2020pccp}. We anticipate that a detailed study of the far richer intramolecular vibrational energy flow pathways along with the invariant manifolds that regulate the phase space transport will lead to control strategies based on exploiting the dynamics in the vicinity of the higher index saddles.

\section*{ACKNOWLEDGMENTS}
The authors would like to acknowledge the fruitful discussions with Stephen Wiggins, the tireless support of Eleanor Machin during the COVID-19 lockdown in Bristol and  IIT Kanpur High Performance Computing facility for computing resources.

\section*{FUNDING}
Priyanka Pandey is supported by a graduate fellowship from IIT Kanpur; Srihari Keshavamurthy's research is supported by the Science and Engineering Research Board (SERB) India (project no. EMR/006246). Shibabrat Naik acknowledges the support of EPSRC Grant No. EP/P021123/1.

\section*{CONFLICT OF INTEREST}

The authors declare that they have no conflicts of interest.

\section*{AUTHORS CONTRIBUTIONS}
Priyanka Pandey performed the classical and quantum dynamical computations. Shibabrat Naik contributed in terms of the computation of the classical invariant manifolds and their intersections.  All authors participated in discussing the results and were involved in writing the text of the paper.

\section*{SUPPLEMENTARY MATERIALS}
Supplementary material with details on computation of the invariant manifolds is available for this article at and are accessible for authorized users.




\begin{thebibliography}{99}


\bibitem{eyring1935activated}Eyring, H., The activated complex in chemical reactions, \textit{J. Chem. Phys.}, 1935, vol.\,3, no.\,2, pp.\,107-115.

\bibitem{maeda2013systematic}Maeda, S., Ohno, K., and Morokuma, K., Systematic exploration of the mechanism of chemical reactions: the global reaction route mapping (GRRM) strategy using the ADDF and AFIR methods, \textit{Phys. Chem. Chem. Phys.}, 2013, vol.\,15, no.\,11, pp.\,3683-3701.

\bibitem{tsutsumi2020visualization}Tsutsumi, T., Ono, Y., Arai, Z., and Taketsugu, T., Visualization of dynamics effect: projection of on-the-fly trajectories to the subspace spanned by the static reaction path network, \textit{J. Chem. Theory Comput.}, 2020, vol.\,16, no.\,7, pp.\,4029–4037. 

\bibitem{pollak_transition_1978}Pollak, E., and Pechukas, P., Transition states, trapped trajectories, and classical bound states embedded in the continuum, \textit{J. Chem. Phys.}, 1978, vol.\,69, no.\,3, pp.\,1218-1226.

\bibitem{pechukas_classical_1979}Pechukas, P., and Pollak, E., Classical transition state theory is exact if the transition state is unique, \textit{J. Chem. Phys.}, 1979, vol.\, 71, no.\,5, pp.\, 2062-2068.

\bibitem{waalkens_wigners_2008} Waalkens, H., Schubert, R., and Wiggins, S., Wigner's dynamical transition state theory in phase space: classical and quantum, \textit{Nonlinearity}, 2008, vol.\,21, no.\,1, pp.\, R1-R118.

\bibitem{uzer2002geometry}Uzer, T., Jaff{\'e}, C., Palaci{\'a}n, J., Yanguas, P., and Wiggins, S., The geometry of reaction dynamics,
\textit{Nonlinearity}, 2002, vol.\,15, no.\,4, pp.\,957.

\bibitem{lu2014evidence} Lu, Z., Chang, Y.\,C., Yin, Q.\,Z., Ng, C.\,Y., and Jackson, W.\,M., Evidence for direct molecular oxygen production in CO2 photodissociation, \textit{Science}, 2014, vol.\,346, no.\,6205, pp.\,61-64.

\bibitem{quapp2015embedding} Quapp, W., and Maria Bofill, J., Embedding of the Saddle Point of Index two on the PES of the Ring Opening of Cyclobutene, \textit{Int. J. Quantum Chem.}, 2015, vol.\,115, no.\,23, pp.\,1635-1649.

\bibitem{breulet1984conrotatory}Breulet, J., and Schaefer III, H.\,F., Conrotatory and disrotatory stationary points for the electrocyclic isomerization of cyclobutene to cis-butadiene, \textit{J. Am. Chem. Soc.}, 1984, vol.\,106, no.\,5, pp.\,1221-1226.

\bibitem{murrell1968symmetries} Murrell, J.\,N., and Laidler, K.\,J., Symmetries of activated complexes,
\textit{Transactions of the Faraday Society}, 1968, vol.\,64, pp.\,371-377.

\bibitem{berry1992limitations} Wales, D.\,J., and Berry, S.\,R.,  Limitations of the Murrell–Laidler theorem, 
\textit{J. Chem. Soc. Faraday Trans.}, 1992, vol.\,88, no.\,4, pp.\,543-544.

\bibitem{heidrich1986saddle} Heidrich, D., and Quapp, W., Saddle points of index 2 on potential energy surfaces and their role in theoretical reactivity investigations, \textit{Theor. Chim. Acta}, 1986, vol.\,70, no.\,2, pp.\,89-98.

\bibitem{stanton1975group} Stanton, R.\,E., and McIver Jr, J.\,W., Group theoretical selection rules for the transition states of chemical reactions, \textit{J. Am. Chem. Soc.}, 1975, vol.\,97, no.\,13, pp.\,3632-3646.

\bibitem{trindle1975group} Trindle, C., and Bouman, T.\,D., Group theory and reaction mechanisms: An extension of the Mciver‐Stanton rules, \textit{Int. J. Quantum Chem.}, 1975, vol.\,9, no.\,S9, pp.\,255-264.

\bibitem{harabuchi2016nontotally} Harabuchi, Y., Ono, Y., Maeda, S., Taketsugu, T., Keipert, K., and Gordon, M.\,S., Nontotally symmetric trifurcation of an S$_{\rm N}$2 reaction pathway, \textit{J. Comput. Chem.}, 2016, vol.\,37, no.\,5, pp.\,487-493.

\bibitem{quapp2018toward} Quapp, W., Bofill, J.\,M., and Ribas‐Ari{\~n}o, J., Toward a theory of mechanochemistry: Simple models from the very beginnings, \textit{Int. J. Quantum Chem.}, 2018, vol.\,118, no.\,23, pp.\,e25775.

\bibitem{haruta2019force}Haruta, N., de Oliveira, P.\,F.\,M., Sato, T., Tanaka, K., and Baron, M., Force-Induced Dissolution of Imaginary Mode in Mechanochemical Reaction: Dibenzophenazine Synthesis, \textit{J. Phys. Chem. C}, 2019, vol.\,123, no.\,35, pp.\,21581-21587.

\bibitem{wollenhaupt2015should}Wollenhaupt, M., Krupi{\v{c}}ka, M., and Marx, D., Should the Woodward–Hoffmann Rules be Applied to Mechanochemical Reactions?, \textit{ChemPhysChem}, 2015, vol.\,16, no.\,8, pp.\,1593-1597.

\bibitem{ribas2012covalent}Ribas-Arino, J., and Marx, D., Covalent mechanochemistry: theoretical concepts and computational tools with applications to molecular nanomechanics, \textit{Chem. Rev.}, 2012, vol.\,112, no.\,10, pp.\,5412-5487.

\bibitem{mauguiere2013bond} Maugui{\`e}re, F.\,A.\,L., Collins, P., Ezra, G.\,S., and Wiggins, S., Bond breaking in a Morse chain under tension: Fragmentation patterns, higher index saddles, and bond healing, \textit{J. Chem. Phys.}, 2013, vol.\,138, no.\,13, pp.\,134118.

\bibitem{collins2011index}Collins, P., Ezra, G.\,S., and Wiggins, S., Index k saddles and dividing surfaces in phase space with applications to isomerization dynamics, \textit{J. Chem. Phys.}, 2011, vol.\,134, no.\,24, pp.\,244105.

\bibitem{haller2010transition} Haller, G., Uzer, T., Palaci{\'a}n, J., Yanguas, P., and Jaff{\'e}, C., Transition states near rank-two saddles: Correlated electron dynamics of helium, \textit{Comm. Nonlinear Sci. Numer. Simulat.}, 2010, vol.\,15, no.\,1, pp.\,48-59.

\bibitem{nagahata2013reactivity} Nagahata, Y., Teramoto, H., Li, C.\,B., Kawai, S., and Komatsuzaki, T., Reactivity boundaries for chemical reactions associated with higher-index and multiple saddles, \textit{Phys. Rev. E}, 2013, vol.\,88, no.\,4, pp.\,042923.

\bibitem{mann2002ab} Mann, D.\,J., and Hase, W.\,L., Ab initio direct dynamics study of cyclopropyl radical ring-opening,
\textit{J. Am. Chem. Soc.}, 2002, vol.\,124, no.\,13, pp.\,3208-3209.

\bibitem{pradhan2019can} Pradhan, R., and Lourderaj, U., Can reactions follow non-traditional second-order saddle pathways avoiding transition states?, \textit{Phys. Chem. Chem. Phys.}, 2019, vol.\,21, no.\,24, pp.\,12837-12842.

\bibitem{yoshikawa2012quantum} Yoshikawa, T., Sugawara, S., Takayanagi, T., Shiga, M., and Tachikawa, M., Quantum tautomerization in porphycene and its isotopomers: Path-integral molecular dynamics simulations, \textit{Chem. Phys.}, 2012, vol.\,394, no.\,1, pp.\,46-51.

\bibitem{smedarchina2018entanglement}Smedarchina, Z., Siebrand, W., and Fern{\'a}ndez-Ramos, A., Entanglement and co-tunneling of two equivalent protons in hydrogen bond pairs, \textit{J. Chem. Phys.}, 2018, vol.\,148, no.\,10, pp.\,102307.

\bibitem{fillaux2005quantum}Fillaux, F., Quantum entanglement and nonlocal proton transfer dynamics in dimers of formic acid and analogues, \textit{Chem. Phys. Lett.}, 2005, vol.\,408, no.\,4-6, pp.\,302-306.

\bibitem{litman2019elucidating}Litman, Y., Richardson, J.\,O., Kumagai, T., and Rossi, M., Elucidating the nuclear quantum dynamics of intramolecular double hydrogen transfer in porphycene, \textit{J. Am. Chem. Soc.}, 2019, vol.\,141, no.\,6, pp.\,2526-2534.

\bibitem{rumpel1989nmr}Rumpel, H., and Limbach, H.\,H., NMR study of kinetic HH/HD/DD isotope, solvent and solid-state effects on the double proton transfer in azophenine, \textit{J. Am. Chem. Soc.}, 1989, vol.\,111, no.\,14, pp.\,5429-5441.

\bibitem{meschede1991dynamic}Meschede, L., and Limbach, H.\,H., Dynamic NMR study of the kinetic HH/HD/DD isotope effects on the double proton transfer in cyclic bis (p-fluorophenyl) formamidine dimers, \textit{J. Phys. Chem.}, 1991, vol.\,95, no.\,25, pp.\,10267-10280.

\bibitem{albery1986isotope}Albery, W.\,J., Isotope effects in double-proton-transfer reactions, \textit{J. Phys. Chem.}, 1986, vol.\,90, no.\,16, pp.\,3774-3783.

\bibitem{accardi2010synchronous} Accardi, A., Barth, I., K{\"u}hn, O., and Manz, J., From synchronous to sequential double proton transfer: Quantum dynamics simulations for the model Porphine, \textit{J. Phys. Chem. A}, 2010, vol.\,114, no.\,42, pp.\,11252-11262.

\bibitem{ushiyama2001successive}Ushiyama, H., and Takatsuka, K., Successive mechanism of double-proton transfer in formic acid dimer: A classical study, \textit{J. Chem. Phys.}, 2001, vol.\,115, no.\,13, pp.\,5903-5912.

\bibitem{walewski2010car}Walewski, L., Waluk, J., and Lesyng, B., Car-Parrinello molecular dynamics study of the intramolecular vibrational mode-sensitive double proton-transfer mechanisms in porphycene, \textit{J. Phys. Chem. A}, 2010, vol.\,114, no.\,6, pp.\,2313-2318.

\bibitem{black2012dynamics}Black, K., Liu, P., Xu, L., Doubleday, C., and Houk, K.\,N., Dynamics, transition states, and timing of bond formation in Diels–Alder reactions, \textit{Proc. Natl. Acad. Sci. U.S.A.}, 2012, vol.\,109, no.\,32, pp.\,12860-12865.

\bibitem{takeuchipnas2007} Takeuchi, S., and Tahara, T., The Answer to Concerted Versus Step-wise controversy for the Double Proton Transfer Mechanism of $7$-Azaindole Dimer in Solutiuon, \textit{Proc. Natl. Acad. Sci. (USA)}, 2007, vol.\,104, pp.\,5285-5290.

\bibitem{smedarchina2007correlated} Smedarchina, Z., Siebrand, W., and Fern{\'a}ndez-Ramos, A., Correlated double-proton transfer. I. Theory, \textit{J. Chem. Phys.}, 2007, vol.\,127, no.\,17, pp.\,174513.

\bibitem{smedarchina2008mechanisms} Smedarchina, Z., Siebrand, W., Fern{\'a}ndez-Ramos, A., and Meana-Paneda, R., Mechanisms of double proton transfer. Theory and applications, \textit{Z. Phys. Chem.}, 2008, vol.\,222, no.\,8, pp.\,1291.

\bibitem{albaredajpcl2015} Albareda, G., Bofill, J.\,M., Tavernelli, I., Huarte-Larranaga, F., Illas, F., and Rubio, A., Conditional Born-Oppenheimer Dynamics: Quantum Dynamics Simulations for the Model Porphine, \textit{J. Phys. Chem. Lett.}, 2015, vol.\,6, pp.\,1529-1535.

\bibitem{tannor2007introduction} Tannor, D.\,J., \textit{Introduction to quantum mechanics: a time-dependent perspective}, University Science Books, 2007.

\bibitem{dion2014program} Dion, C.\,M., Hashemloo, A., and Rahali, G., Program for quantum wave-packet dynamics with time-dependent potentials, \textit{Comput. Phys. Commun.}, 2014, vol.\,185, no.\,1, pp.\,407-414.

\bibitem{blanes2006symplectic} Blanes, S., Casas, F., and Murua, A., Symplectic splitting operator methods for the time-dependent Schr{\"{o}}dinger equation, \textit{J. Chem. Phys.}, 2006, vol.\,124, no.\,23, pp.\,234105.

\bibitem{bandrauk1991improved} Bandrauk, A.\,D., and Shen, H., Improved exponential split operator method for solving the time-dependent Schr{\"o}dinger equation, \textit{Chem. Phys. Lett.}, 1991, vol.\,176, no.\,5, pp.\,428-432.

\bibitem{li2014efficient}Li, W., Zhang, D.\,H., and Sun, Z., Efficient fourth-order split operator for solving the triatomic reactive schrodinger equation in the time-dependent wavepacket approach, \textit{J. Phys. Chem. A}, 2014, vol.\,118, no.\,42, pp.\,9801-9810.

\bibitem{helmkamp1994structures} Helmkamp, B.\,S., and Browne, D.\,A., Structures in classical phase space and quantum chaotic dynamics,
\textit{Phys. Rev. E}, 1994, vol.\,49, no.\,3, pp.\,1831.

\bibitem{vela2001time} Vela-Arevalo, L.\,V., and Wiggins, S., Time-frequency analysis of classical trajectories of polyatomic molecules, \textit{Int. J. Bifurcation Chaos Appl. Sci. Eng.}, 2001, vol.\,11, no.\,05, pp.\,1359-1380.

\bibitem{ksacp2013} Keshavamurthy, S.,  Scaling Perspective on Intramolecular Vibrational Energy Flow: Analogies, Insights, and Challenges, \textit{Adv. Chem. Phys.}, 2013, vol.\, 153, pp.\,43-110.

\bibitem{wiggins_role_2016} Wiggins, S., The role of normally hyperbolic invariant manifolds (NHIMS) in the context of the phase space setting for chemical reaction dynamics, \textit{Regul. Chaotic Dyn.}, 2016, vol.\, 21, no.\, 6, pp.\,621-638.

\bibitem{marston1989reactive} Marston, C.\,C., and De Leon, N., Reactive islands as essential mediators of unimolecular conformational isomerization: A dynamical study of 3‐phospholene, \textit{J. Chem. Phys.}, 1989, vol.\,91, no.\,6,  pp.\,3392-3404.

\bibitem{husimi1940some} Husimi, K., Some formal properties of the density matrix,in \textit{Proc. Phys. Math. Sot. Japan. 3rd Series}, 1940, vol.\,22, no.\,4, pp.\,264-314.

\bibitem{wiggins1990geometry} Wiggins, S., On the geometry of transport in phase space I. Transport in k-degree-of-freedom Hamiltonian systems, $2 
\leqslant k < \infty$, \textit{Physica D: Nonlinear Phenomena}, 1990, vol.\, 44, no.\, 3, pp.\,471-501.

\bibitem{naik2017computational} Naik, S., Lekien, F., and Ross, S. D., Computational method for phase space transport with applications to lobe dynamics and rate of escape. \textit{Regul. Chaotic Dyn.}, 2017, vol.\, 22, no.\, 3, pp.\,272-297.

\bibitem{mendoza2010hidden} Mendoza, C., and Mancho, A.\,M., Hidden geometry of ocean flows, \textit{Phys. Rev. Lett.}, 2010, vol.\,105, no.\,3, pp.\,038501.

\bibitem{patra2018detecting} Patra, S., and Keshavamurthy, S., Detecting reactive islands using Lagrangian descriptors and the relevance to transition path sampling, \textit{Phys. Chem. Chem. Phys.}, 2018, vol.\,20, no.\,7, pp.\,4970-4981.

\bibitem{mancho2013lagrangian} Mancho, A.\,M., Wiggins, S., Curbelo, J., and Mendoza, C., Lagrangian descriptors: A method for revealing phase space structures of general time dependent dynamical systems, \textit{Comm. Nonlinear Sci. Numer. Simulat.}, 2013, vol.\,18, no.\,12, pp.\,3530-3557.

\bibitem{garcia2020exploring} Garc{\'\i}a-Garrido, V.\,J., Agaoglou, M., and Wiggins, S., Exploring Isomerization Dynamics on a Potential Energy Surface with an Index-2 Saddle using Lagrangian Descriptors, \textit{Comm. Nonlinear Sci. Numer. Simulat.}, 2020, vol.\,89, pp.\,105331.

\bibitem{nagahata2020pccp} Nagahata, Y., Borondo, F., Benito, R. M., and Hernandez, R., Identifying Reaction Pathways in Phase Space via Asymptotic Trajectories, \textit{Phys. Chem. Chem. Phys.}, 2020, vol.\,22, pp.\,10087-10105.

\end{thebibliography}

\endpaper

\end{document}